\documentclass[%
preprint,
 amsmath,amssymb,
 aps,
]{revtex4-2}

\usepackage{graphicx}
\usepackage{dcolumn}
\usepackage{bm}
\usepackage[mathlines]{lineno}

\begin{document}

\preprint{APS/123-QED}

\title{Vorticity Suppression by Particle Lag Effects in Shock-Driven Multiphase Instability}

\author{Vasco O. Duke-Walker}
\author{Jacob A. McFarland}%
 \email{mcfarlandja@tamu.edu}
\affiliation{Texas A\&M University}%




\date{\today}

\begin{abstract}

Shock-driven multiphase mixing occurs in many physical systems such as explosive dispersal of chemical or biological agents, in the evolution of supernova remnants, and in supersonic and detonative combustion engines. This mixing process is driven by the Shock Driven Multiphase Instability (SDMI), a derivative of the canonical Richtmyer-Meshkov Instability (RMI). The SDMI deviates from the RMI as particle lag effects become significant, where a higher momentum deficit leads to longer equilibration times and a reduction in hydrodynamic mixing.

In this work, the effect of particle lag (rate of momentum transfer) on the SDMI evolution was isolated and investigated utilizing solid nondeforming and nonevaporating particles of differing sizes while holding the effective density ratio (mass of particles in the interface) constant. Three particle sizes were selected with increasing velocity relaxation times. Experiments were conducted by accelerating a cylindrical interface comprised of air and seeded particles surrounded by clean (particle-free) air with a Mach $\sim1.35$ shock wave. The development of the multiphase interface was measured using particle imaging velocimetry (PIV). Circulation measurements showed a decrease in mixing with increasing particle size. Finally, a new model, derived from theory, is proposed to predict circulation deposition, mixing energy, in the SDMI based on shock strength, effective density ratio, and particle response times.

\end{abstract}

\maketitle


\section{\label{sec:Introduction}Introduction and Previous Work}

\begin{figure}[ht]
	\centering
		\includegraphics[width=1\textwidth]{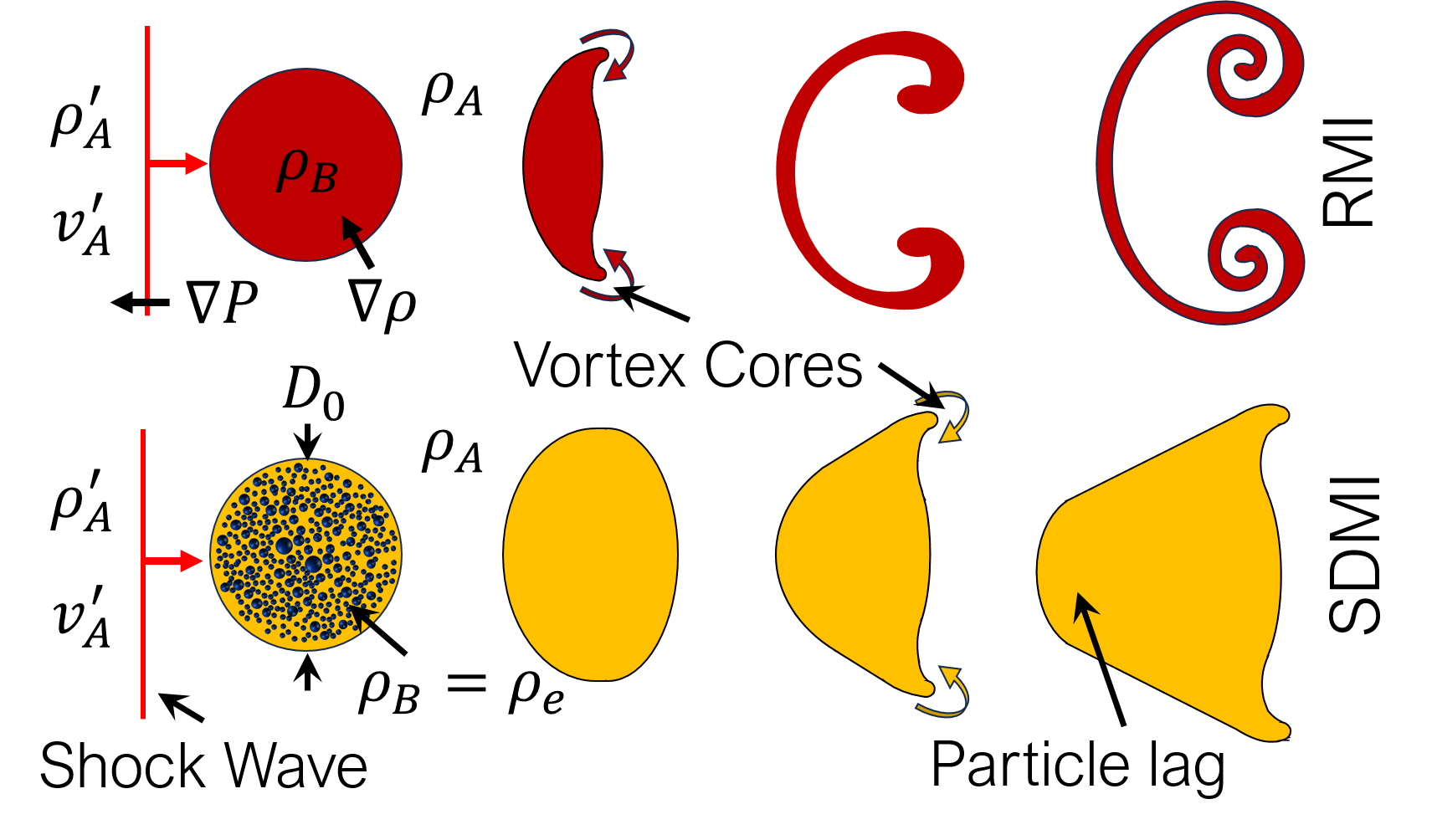}
	  \caption{Evolution of the Shock-Driven Multiphase Instability compared to the Richtmyer-Meshkov Instability.}
	  \label{figure:PL_Evolution}
\end{figure}

The Shock-Driven Multiphase Instability (SDMI) develops when an interface between multiphase fluids is impulsively accelerated by a shock wave. The multiphase fluid is composed of a fluid (gas here) carrier phase and a dispersed particle phase consisting here of spherical non-deforming and non-evaporating solid particles. During the SDMI's development, the shock wave instantaneously accelerates the carrier gas, while the acceleration of the particles is delayed, depending on their respective mass and drag force. Smaller/lighter particles will rapidly accelerate to equilibrium, while larger/heavier particles will lag significantly, exchanging momentum with the gas over a finite equilibration time (Fig. \ref{figure:PL_Evolution}). In the small-particle limit (infinitesimally small particles) momentum coupling is nearly instantaneous, and the SDMI behaves similar to the single fluid Richtmyer-Meshkov Instability (RMI). 

The RMI is a fluid instability characterized by the development of vorticity ($\omega_k$) due to a misalignment between pressure and density gradients. This misalignment facilitates the deposition of baroclinic vorticity along the interface of fluids with varying densities as can be seen in the vorticity eqn \ref{eqn:vort}, where $u_i$ is the gas velocity, $\rho_g$ in the gas density, $\tau_{jl}$ is the viscous stress tensor, and $B_j$ is a body force. The baroclinic source term (third term on right hand side) deposits strong vorticity along the interface, inducing growth and stretching of the fluid interface. This vorticity fosters mixing of the different species across the interface and the generation of multiple mixing length scales. This dynamic process ultimately leads to turbulent-like mixing. 

\begin{eqnarray}
  \partial_t \omega_k + u_j \partial_j \omega_i =& \omega_j \partial_j u_i - \omega_i \partial_j u_j +  \frac{1}{\rho_g^2} \varepsilon_{kij} \partial_i \rho_g \partial_j P + \varepsilon_{kij} \partial_i \frac{\partial_l \tau_{jl}}{\rho_g} + \varepsilon_{kij} \partial_i \frac{B_j}{\rho_g}
\label{eqn:vort}.
\end{eqnarray}

The total vorticity deposited on an interface can be measured by the circulation (eqn \ref{eqn:Gamma}). Several models have been proposed to estimate the circulation deposition on a circular RMI interface. Wang et al. \cite{wang2015interaction} provided a summary and comparison of various models to simulation results. Yang et al. \cite{yang1994model} proposed a model for a light gas cylinder while Picone and Boris \cite{picone1988vorticity} proposed a model with better accuracy for a heavy gas cylinder. This model, of particular interest to this work, is shown in eqn. \ref{eqn:PBcirc}, where $v_A'$ is the shocked surrounding gas velocity, $w_i$ the incident shock wave velocity, $D$ the interface diameter, ,and $\rho_{A}$ and $\rho_{B}$  are the surrounding and interface gases (see Fig. \ref{figure:PL_Evolution}).

\begin{equation}
    \Gamma = \iint \omega_{i,j} dA 
\label{eqn:Gamma}
\end{equation}

\begin{equation}
    \label{eqn:PBcirc}
    \Gamma_PB=v_{A}'\big(\frac{v_{A}'}{w_i}-2\big)\frac{D}{2}*ln(\rho_B/\rho_A)
\end{equation}


For a particle-laden gas, vorticity is generated from the body force source term that results from momentum transfer through the particle drag force (last term on the right hand side). It was observed by Dahal and McFarland \cite{dahal2017numerical} that vorticity decreased as particle size, and lag distance, increased. The vorticity equation contains no term to describe the reduction in vorticity deposition due to the misalignment of the particle source term, as all terms are assumed to advect with the gas flow. Dahal and McFarland instead proposed the use of the enstrophy, eqn. \ref{eqn:enst}, where $\Omega=\omega_k \omega_k/2$. In this equation, the enstrophy production can be seen to increase when the enstrophy source term and previously deposited vorticity are aligned (last term on the right-hand side). 

\begin{eqnarray}
  D_t \Omega =& \omega_i \omega_j \partial_j u_i - \omega_i \omega_i \partial_j u_j + \omega_k \frac{1}{\rho_g^2} \varepsilon_{kij} \partial_i \rho_g \partial_j P +  \nu \partial_j \partial_j \Omega -\nu \partial_j \omega_i \partial_j \omega_i + \omega_k \varepsilon_{kij} \partial_i \frac{B_j}{\rho_g}
\label{eqn:enst}.
\end{eqnarray}

As particle size increases, the source term lags further behind the flow and the SDMI departs from the classic RMI instability. The SDMI depends on three main parameters: the shock strength, effective Atwood number, and particle relaxation time. The shock strength and effective Atwood number drive the potential for mixing, as in the RMI. On the other hand, increasing the particle relaxation time suppresses mixing.

The pressure gradient can be derived from gas dynamics for a given shock strength and gas properties while the density gradient is described by the Atwood number for gas or gas-particle systems \cite{mcfarland2016computational}. In the case of a gas density gradient the gas Atwood number can be defined as $A_g= \frac{\rho_{B} -\rho_{A}}{\rho_{B} +\rho_{A}}$. For a gas/gas-particle interface the gradient of density can be characterized by the effective Atwood number as $A_{e}= \frac{\rho_{e} -\rho_{A}}{\rho_{e} +\rho_{A}}$ \cite{ukai2010richtmyer, vorobieff2011vortex}, where $\rho_{e}$ is the effective density of the mixture (combined mass of each phase divided by the total volume). The relative effect of the finite particle velocity equilibration time can be estimated from the ratio of the velocity relaxation time $t_v$ to the characteristic hydrodynamic time $t_c$, $\tau_v = t_v/t_c$. A simple estimate of the hydrodynamic time for a shock-driven circular interface can be made as $t_c = D/w_i$. The velocity relaxation time is a function of the particle mass and drag force history.

For solid spherical particles, complex effects such as deformation, breakup, mass transfer, and reaction can be neglected and the particle drag force can be found from well-developed drag relations. Various forces arise on a spherical particle as it is accelerated \cite{maxon2021high,parmar2010improved}, but among these the steady drag force is the longest lasting and most influential. The drag force depends on the Reynolds number ($Re=\rho_B' d_p (v_A'-v_p)/\mu_B'$), where $\rho_B'$ is the post shock carrier gas density, $d_p$ and $v_p$ are the particle diameter and velocity, $\mu_B'$ the post-shock carrier gas viscosity, and $v_A'$ the carrier gas post-shock velocity. Here it is assumed that $v_A'\approx v_B'$ as the interface gas volume is limited and will equilibrate with the surrounding gas velocity rapidly. 

For low Reynolds numbers, $Re\lessapprox 1$, the Stokes flow drag relation can be used, where the drag coefficient is calculated as $C_D=24/Re$. It is assumed here that equilibrium is achieved when the final velocity of the particle reaches 99\% of the freestream gas velocity, $v_{p,f}=0.99v_A'$, transferring ~98\% of its momentum to the gas phase. From the drag force, the particle velocity relaxation time can be defined as $t_{v,s}=-ln(0.01)/B$, where $B=18\mu_B'/(d_p^2\rho_p)$and $\rho_p$ is the density of the particle material. The distance the particle travels relative to the gas at this time, the total lag distance ($x_f$), can be found as shown in eqn \ref{eqn:xlag}. 

\begin{equation}
    \label{eqn:xlags}
    \begin{aligned}
    x_s=\frac{0.99u_0}{B}
    \end{aligned}
\end{equation}

At higher $Re$, Cloutman \cite{cloutman1988analytical} provides an analytical solution for particle trajectories based on the Kliatchko drag model (eqn. \ref{eqn:drag}. For the particle sizes used in this work, $Re<1000$ and thus, only this solution is presented. Cloutman also provides another solution for $Re>1000$. The high $Re$ velocity relaxation time can be found as shown in eqn. \ref{eqn:tv}, where $C=\frac{1}{6}[d_p \rho_g/(\mu)]^{2/3}$. The lag distance, can be then estimated utilizing eqn. \ref{eqn:xlag}.

\begin{equation}
\label{eqn:drag}
  C_{D} = \left\{
    \begin{array}{ll}
        24/Re+4/{{Re}}^{1/3} &  Re \le 1000  \\ 
        0.424         & {{Re}} > 1000
    \end{array} \right.
\end{equation}

\begin{equation}
    \label{eqn:tv} 
    t_{v,c}=\frac{3}{2B} ln\left(\frac{v_{p,f}^{-2/3}+C}{v_A'^{-2/3}+C}\right)
\end{equation}

\begin{equation}
    \label{eqn:xlag}
    \begin{aligned}
    x_c=\frac{3}{BC}\bigg(&v_A'^{\frac{1}{3}} +C^{-\frac{1}{2}}\arctan{C^{-\frac{1}{2}}v_A'^{-\frac{1}{3}}}\\ -&v_{p,f}^{\frac{1}{3}}-C^{-\frac{1}{2}}\arctan{C^{-\frac{1}{2}}v_{p,f}^{-\frac{1}{3}}}\bigg)
    \end{aligned}
\end{equation}

Several investigations have been conducted through simulation and experimentation to explore the effect of particle lag (velocity relaxation time) on shock-driven multiphase mixing. Simulation efforts  \cite{dahal2017numerical, mcfarland2016computational,black2017evaporation,black2018particle} showed that the vorticity deposition and mixing decrease as particle relaxation times increase. Thus, large particles resulted in a lower interface development and mixing. Dahal and McFarland \cite{dahal2017numerical} utilized the enstrophy equation to explain that vorticity competition from the advecting particle source terms was responsible for the reduced vorticity in the case of large particles. In addition, the paper highlighted the importance of the particle size distribution when estimating the development of the hydrodynamics instabilities in the flow. Black et al. \cite{black2018particle} presented a comparison between the classic RMI (dusty gas case) and three different particle size cases. It was noted that the strength of the vorticity reduces with increasing particle size. In addition, heating effects from the particle drag were noted to further increase dissipation of circulation. Following this work, Paudel et al. \cite{paudel2018particle} extended it utilizing 3D simulations and showed the importance of the spatial distribution of the particles to the prediction of the hydrodynamics times scales, ultimately modifying the mass transfer during evaporation. 

Experimentally it has been qualitatively shown that the classical RMI development is damped or diminished due to the particle response time, mainly due to drag forces \cite{vorobieff2011vortex,  vorobieffl2013morphology,middlebrooks2018droplet}. Previous studies (\cite{vorobieff2011vortex, vorobieffl2013morphology}) identified the uniqueness of the SDMI from the RMI, but the lack of quantitative measurements made it difficult to distinguish what portion of the mixing was primarily driven by the particles. Following this work, Middlebrooks et al. \cite{middlebrooks2018droplet} showed the development of a cylindrical multiphase interface accelerated by a a Mach 1.66 incident shock wave. This work used two different water droplet mean sizes ($d_p \sim 2 [\mu m]$ and $d_p \sim 11 [\mu m]$) and two different effective Atwood numbers ($A_{eff} \sim 0.014 $ and $A_{eff} \sim 0.07 $) to highlight the effect of particle relaxation time and instability strength. It demonstrated that large particles would damp the mixing between the particle-seeded gas and the unseeded gas. However, it would also increase particle entrainment into the unseeded gas as the largest particles fell behind the mixing interface. While this work showed promising results, secondary phenomena such as droplet breakup and evaporation complicated the problem, obscuring the effect of particle lag on cloud mixing. 

The effect of evaporation \cite{duke2020method} and droplet breakup \cite{duke2023experiments,duke2021evaporation} were explored separately in further work. Duke-Walker et al.\cite{duke2020method} showed that breakup enhanced the evaporation process early on, however at later times hydrodynamic mixing limited evaporation rates. A combined simulation and experimental work \cite{ duke2021evaporation} provided greater insight into the results of Middlebrooks et al. and validated particles models for SDMI. Further work from Duke Walker et al.\cite{duke2023experiments} isolated the droplet scale physics from hydrodynamic mixing and studied the droplet breakup and evaporation rates, providing a model for child droplet cloud mixing and evaporation. 

The current work intends to deepen our understanding of the SDMI and multiphase mixing by leveraging solid particles to isolate particle lag effects from breakup and evaporation effects, allowing a deeper understanding of momentum transfer and multiphase mixing. To do so, three different  particle types, with differing velocity relaxation times, were explored by varying particle size, and material properties, maintaining the a constant effective Atwood number ($A_{eff}\sim 0.04$). Small fast-reacting particles ($\tau_v< 0.1$), medium transitional ($1<\tau_v >10$), and large slow reacting particles ($\tau_v >10$ ) were used. From the experimental results, vorticity deposition was quantified by circulation and compared with theoretical predictions for the RMI \cite{picone1988vorticity}. A new model for vorticity deposition by advecting particles was derived from the vorticity and enstrophy equations. From this, a circulation model for the SDMI, accounting for particle lag effects, is presented and compared to the current experimental data and previous simulation and experimental data.

\section{\label{sec:Experimental Facility}Experimental Facility}
This section introduces the equipment used for the experiments, including the diagnostics, data acquisition system, and particle-gas interface shaping device. 

\begin{figure*}[ht]
	\centering
	\includegraphics[width=1\textwidth]{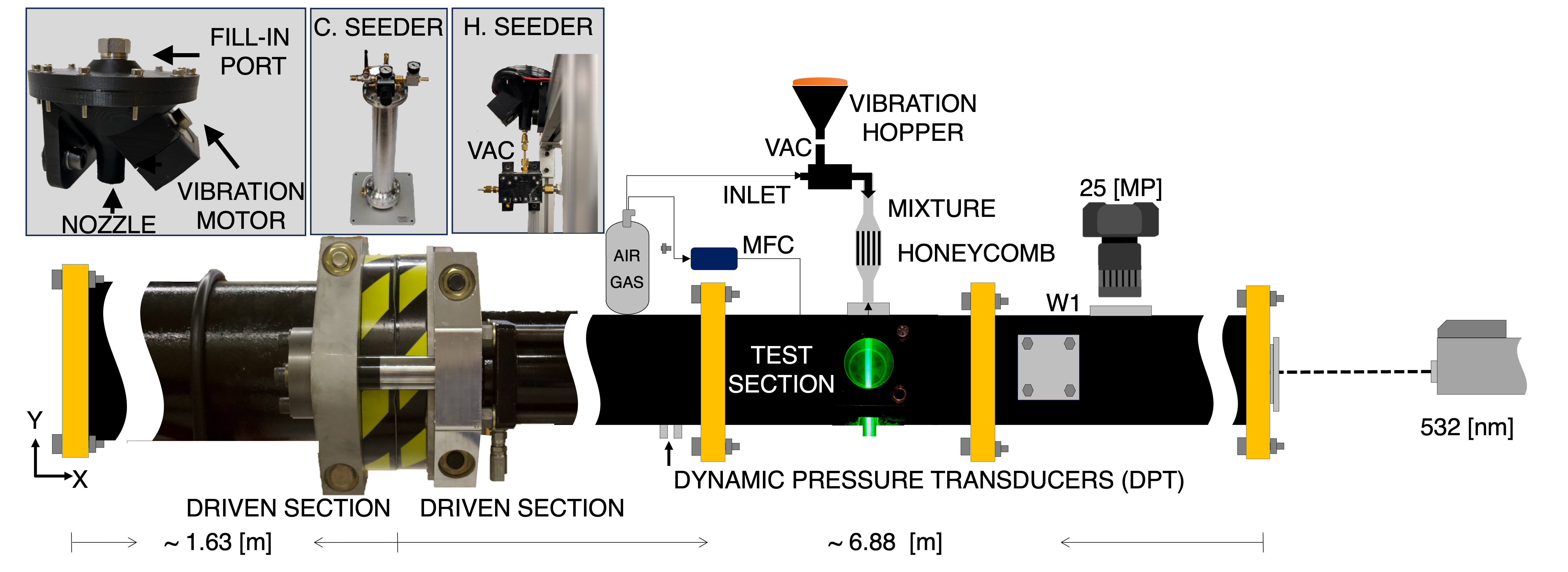}
	  \caption{Shocktube experimental facility}
	  \label{figure:Facility}
\end{figure*}

The experiments investigating fluid mixing were carried out in a shock tube facility, shown in figure \ref{figure:Facility}. The shock tube consists of three primary sections: the driver, driven, and test section, which houses the multiphase interface and diagnostics. A diaphragm, made out of clear poly-carbonate with a thickness of approximately $254 \pm 25.4$ [$\mu m$], is placed between the driver and driven sections and then clamped into place using two 50 kip dual actuating hydraulic rams to seal the system. The driver section is pressurized to $24 \pm 1 [psig]$, just below the diaphragm's breaking pressure of $\sim 30 [psig]$. The driver pressure is monitored by a static transducer with a maximum pressure range of $35 \pm 0.125  [psig]$. Then the experiment is initiated by a high-pressure gas pulse from a gas tank held at 850 [psig]. This ruptures the diaphragm with the help of a sharp x-shaped knife nearly instantaneously. This method provides highly reliable and repeatable experiments. The driven section is long enough for a stable planar shock to develop fully before reaching the test section, which is kept at atmospheric conditions. The test section has multiple acrylic windows to observe the particle-seeded field from the sides and above. A laser beam enters the test section through an transparent sapphire or quartz window positioned in the end wall for diagnostic purposes. 

The experimental data acquisition system measures signals at a dynamic rate of 1.25 [MHz], which is synchronized through a LabVIEW program that controls the automatic shock firing sequence. Two dynamic pressure transducers (PCB 113B24-DPTs) were employed to measure the shock velocity, recording the pressure jump times and providing timing signals to the diagnostics in the test section. After the shock passes and initiates the trigger signal from the pressure transducer, the programmable timing unit initiates the laser pulses and synchronizes camera imaging at a precise time. The Litron NanoPIV 200 laser is used for the experiment, delivering $\sim 200 [mJ]$  of laser energy at a wavelength of 532 [nm]. The laser output is first focused on the center of the window location. Then, a laser sheet is formed using a plano-cylindrical convex lens to enable planar-imaging through any of the window ports in the test section, as shown in figure \ref{figure:Facility} and \ref{figure:IC_laser}. The laser thickness can be estimated to be $\sim 1-2 [mm]$ at the focused point, allowing a small region for particles to scatter light. Two 25 [MP] cameras, with a pixel size of $2.5  [\mu m]$, and a 50 [mm] lens, were used at two locations in the test section to capture the initial conditions and the morphology of the particles within the evolved interface. Additional information about the shock tube facility can be found in \cite{duke2023experiments,duke2020method}.

The dispersion of nano/micro-particle powders poses a considerable challenge, as highlighted by Lu et al. \cite{lu2018gravitational}, since small particles agglomerate reducing the flowability of the powder. Kumar et al. \cite{kumar2003preliminary} demonstrated that particles with sizes $\leq$ 100 [$\mu m$] struggle to flow with gravity or under pressure. Instead, as particle size decreased they required vibration-assisted discharge, attributed to the increasing strength of cohesive van der Waals forces leading to particle agglomeration. Another significant challenge in managing fine powder is the need to maintain moisture consistency, as emphasized by \cite{moghimian2021metal}. The presence of moisture can exacerbate agglomeration, making it even more difficult to handle. With this in mind, a systematic strategy has been implemented to ensure experimental consistency, especially during the handling and dispersion of the finest powders. The process involves mixture and sonication with ethanol at 40 [kHz] to prevent clumping. Subsequently, the particles undergo drying in an oven for 4-5 [h] within the temperature range of 80-105 [C], following previous recommendations  \cite{kumar2003preliminary,lu2018gravitational}. The dried particles are stored in a vacuum desiccator to shield them from moisture, thereby extending their usable shelf life. Prior to each experiment, the particles undergo heating and sieving to eliminate agglomeration and maintain optimal flowability. This systematic approach aims to address the challenges associated with particle size, cohesive forces, and moisture consistency during the handling and dispersion of fine dry nano-particle powders.

\begin{figure}[ht]
	\centering
\includegraphics[width=1\textwidth]{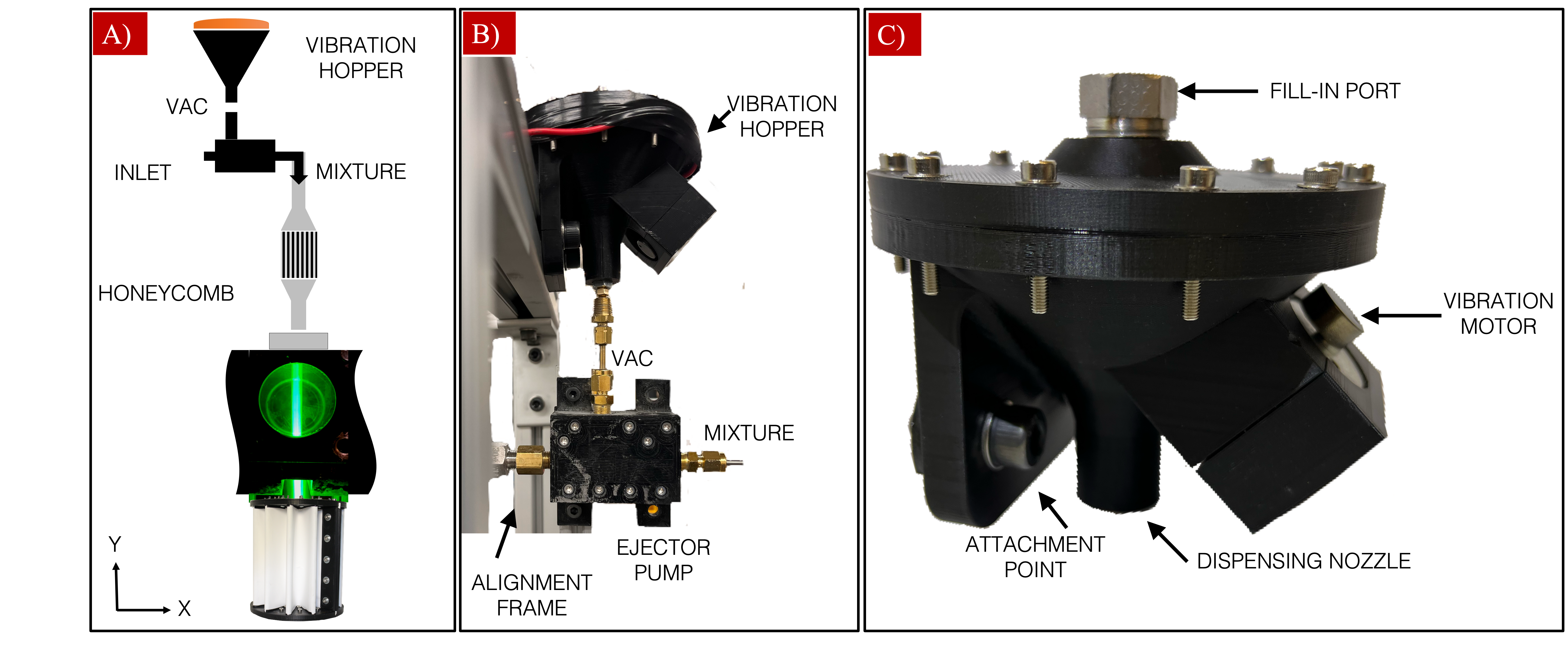}
	  \caption{Particle dispersion system: A) particle mass collection system, B) inline ejector pump and C) vibration hopper}
	  \label{figure:Vhopper}
\end{figure}

\begin{figure}[ht]
	\centering
		\includegraphics[width=0.8\textwidth]{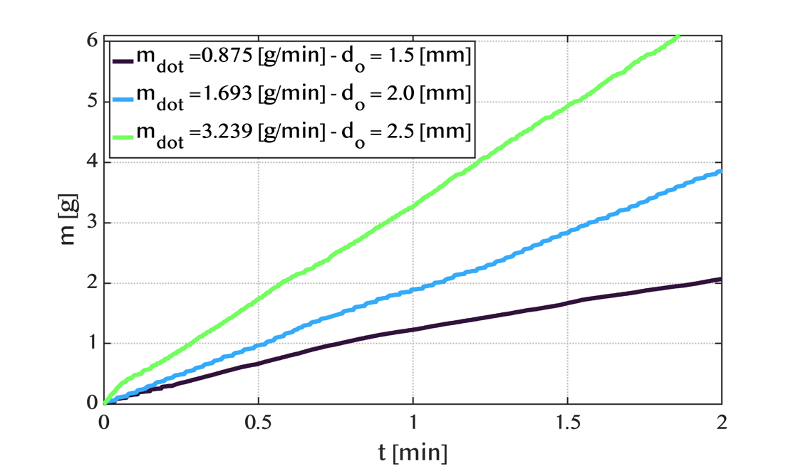}
	  \caption{Aluminium mass discharged example with vibration hopper, utilizing different orifice diameters.}
	  \label{figure:mass_discharged}
\end{figure}

To disperse particle material and create the multiphase interface, an in-house gravity-assisted vibrating hopper system was designed to improve seeding conditions, achieving higher particle concentrations than conventional fluidized-bed seeding devices utilized for commercial particle imaging velocimetry (PIV) systems. As can be seen in figure \ref{figure:Vhopper}, the system is comprised of four main components: 1) a hopper, 2) an interchangeable nozzle exit, 3) a vibration motor, and 4) an ejector pump. The hopper is designed with a 45 [degree] angle \cite{amoros2000design}, allowing a faster feeding rate. Various nozzle sizes are used at the exit of the hopper to control the discharge rate and maintain the highest revolution for a constant seeding condition. The hopper is equipped with a 12[V] 7500[RMP] electric motor with an eccentric mass to assist in discharge and prevent blockages. The ejector pump reduces agglomeration and improves the mixedness of the discharged material with gas flow, ensuring even distribution and facilitating the dispersion process. The particle discharge rate can be controlled with the motor voltage, weight of the eccentric mass, and the nozzle exit diameter. To maintain a steady discharge rate, it is optimal to keep the motor speed high and vary the weight of the eccentric mass to control the amplitude of oscillation in the hopper. Finally, the bigger the nozzle exit diameter, the higher the discharge rate; this can be seen in figure \ref{figure:mass_discharged}. This system provides a high degree of freedom to control the concentration of the multiphase interface. 
   
To create the multiphase interface, the particle seeding system requires the gravity-assisted vibrating hopper and ejector pump, a flow stabilization section, and an external frame. The hopper controls the mass delivered, the ejector pump transports and reduces particle agglomeration, and the flow stabilization minimizes the turbulence to stabilize the interface. The vibration hopper is connected to the test section via an external vibration-isolated frame designed to minimize vibrations at the vacuum inlet that are transmitted throughout the interface. The hopper is then aligned with the vacuum line of the ejector pump, in which air (as carrier gas) was used to transport the mixture into an interface, to distribute the mixture into the flow stabilization section. The ejector pump is located 1 [mm] away from the exit of the vibration hopper and additional atmospheric air in entrained with the particles through this gap. The ejector pump has an orifice diameter of 0.05 [in] that generates a total flow rate of 5.75 [SLPM] between the inlet and vacuum line with an inlet pressure of 6 [Psig]. 

The ejector inlets and outlets are connected through a 2.2 [mm] inner tubing diameter to maintain flow velocity and facilitate smooth particle transition into the stabilization section. The flow is stabilized and shaped by expanding it into a honeycomb and then contracting into a straight 12 [in] 3/8 pipe to form a fully developed laminar jet. In addition, to protect and further stabilize the multiphase jet an annular co-flow was utilized. Lastly, the steady state operation of the hopper was required to deliver a constant mass discharge rate. To confirm this, the vibration hopper was first filled with the desired material then the mass output was collected in steps of 30 seconds. The mass collected was measured with a high precision balance, with an accuracy of 0.1 [mg]. Once the delivery rate reached a steady state (low variation of the mass output), the system was considered to deliver a constant rate of mass. The particle dispersion system was found to achieve a steady state within two minutes of operation.

\section{Experimental Methodology}\label{Experimental Methodology}

This section introduces the experimental procedure executed to explore how the particle response time modifies the SDMI. It covers the description of the initial conditions and the main considerations to guarantee high-fidelity experimental results. 

\subsubsection{Experimental sensitivity analysis} \label{sec:Exp_sens_analysis}

The primary objective of this study is to investigate key variables influencing mixing effects in the SDMI. These variables include the pressure gradient (linked to shock strength), density gradient (described by the effective Atwood number), and the non-dimensional velocity relaxation time (determined by particle size, and density). The pressure and density gradients ultimately govern the maximum strength of vorticity deposition. Reducing the pressure gradient requires a compensatory increase in the density gradient to maintain consistent vorticity deposition (mixing energy). Considering these factors, to achieve similar mixing energy in SDMI development, as in Middlebrooks et al. \cite{middlebrooks2018droplet} Case 2 (t=2700 $\mu s$ $Ae \sim 0.02$ and  $M \sim 1.66 $), at a low shock strength of Mach $\sim 1.35$, the density gradient of the source was increased to $Ae \sim 0.04$. 

The primary motivation behind reducing the shock strength is to maximize the experimental repeatability and accuracy, facilitated by the reliable breakage of diaphragms at lower pressure. Additionally, lower shock strengths/velocities result in lower $Re$, allowing a comparison to common low $Re$ models (i.e. the Stokes drag model) for estimating the particle response time. This strategy aims to streamline experimental procedures and ensure repeatable and precise experiments. 

The effective Atwood number is sensitive to the density of the carrier and surrounding gas (therefore temperature for both cases) and the concentration of the seeded particles. As shown in figure \ref{figure:A_eff} even minor temperature changes can significantly impact the estimations of both the effective and gas Atwood numbers. For this reason, the initial temperatures were monitored constantly just before the experiment was executed to account for the run-to-run variations of conditions in the interface and surrounding gas. Most importantly, to isolate and differentiate the SDMI from the RMI, the gas Atwood number ($A_g$) was set to approximately $\sim 0$. This ensures that the density gradient effect solely arises from particle mass concentration, not from the gas. In doing so, the effective Atwood number ($A_e$) is maintained at a constant value, allowing exploration of the influence of the particle response time in the instability. This is accomplished by controlling the particle and gas mass flow rates from the particle dispensing system.          
         
\begin{figure}[ht]
	\centering
		\includegraphics[width=0.8\textwidth]{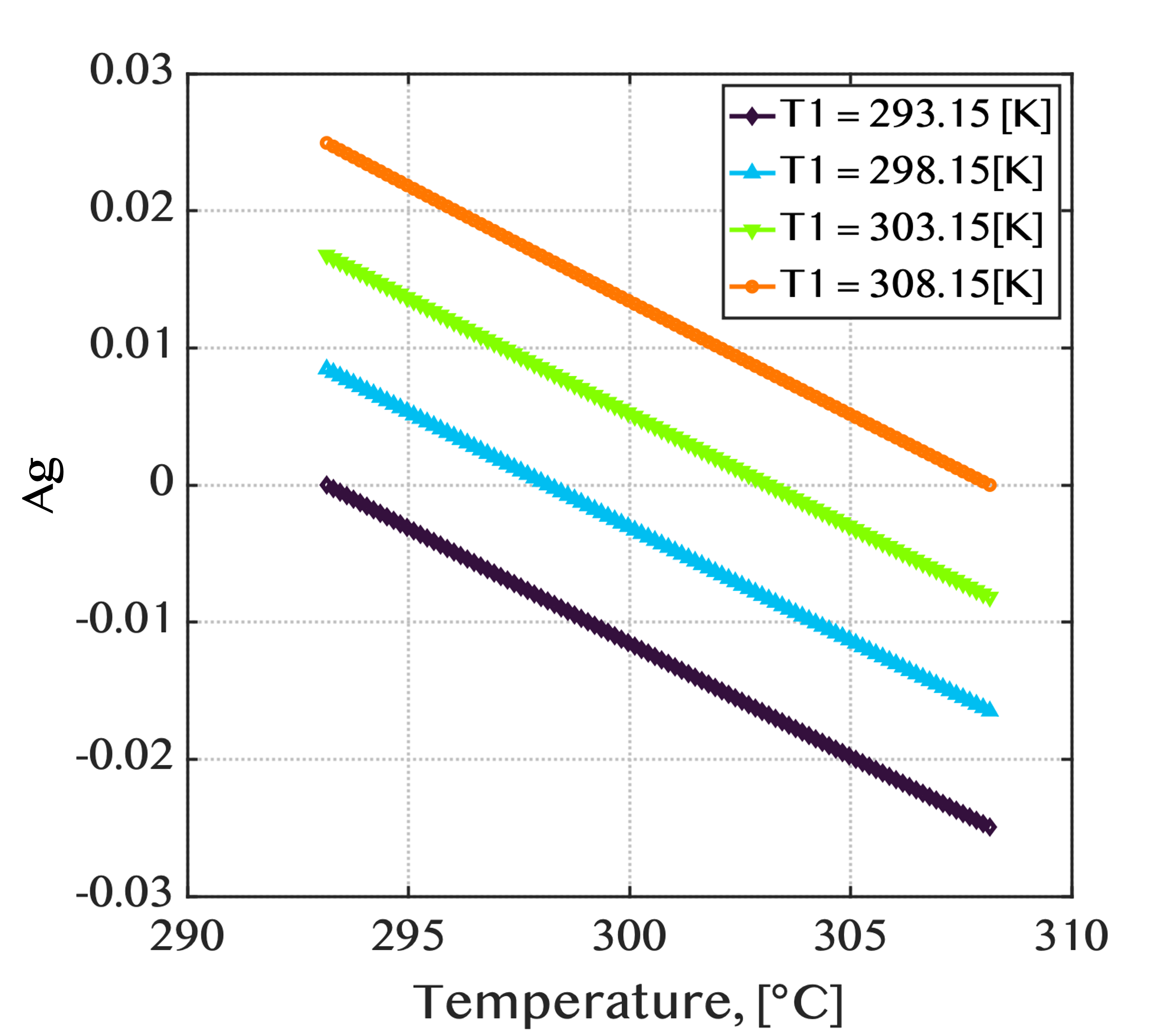}
	  \caption{Atwood number sensitivity to gas temperatures}
	  \label{figure:At_Temp}
\end{figure}

Lastly, to explore the effect of the particle response time on the mixing rate generated, the non-dimensional particle relaxation time must be controlled and prescribed to different values. In previous work \cite{mcfarland2016computational} $t_v$ was estimated utilizing Stokes flows as ${t_v,s}$. However, as the Reynolds number increases, it is important to consider the changes in the drag forces applied to the particle. An analytical solution to the particle motion \cite{cloutman1988analytical} was used to estimate the high-$Re$ velocity relaxation time, $t{_v,c}$, which included the effects of finite Reynolds numbers and varying drag coefficients. While the $t_{v,c}$ provides the superior estimate of the particle lag distance ($x_c$), the Stokes formulation (yielding $x_s$) is found to provide a useful solution for lag effects on vorticity deposition later (see \S \ref{sec:modeling}). Thus, both non-dimensional relaxation times ($\tau_s$ and $\tau_c$) are provided for reference, but $\tau_c$ is primarily used to describe the regimes of particle lag effects. 

\begin{figure}
    \centering
    \includegraphics[width=1\textwidth]{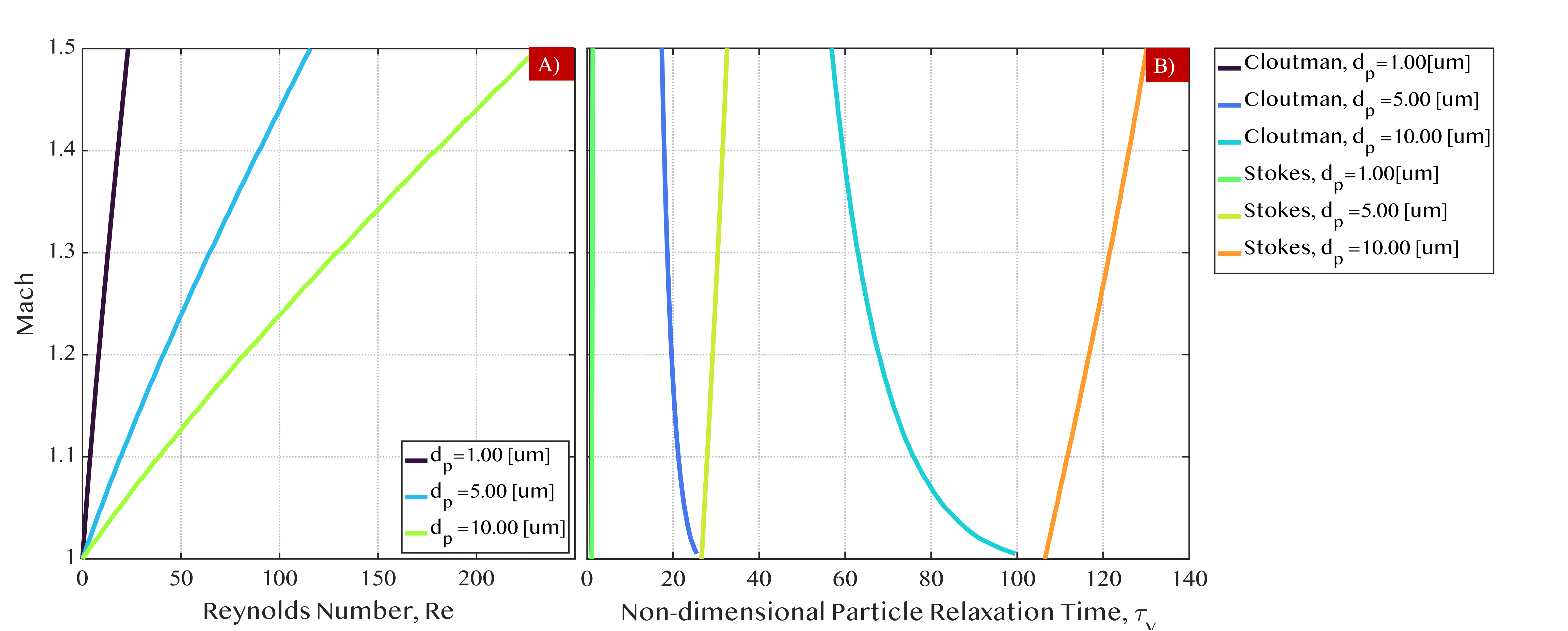}
    \caption{Non-dimensional Reynolds number and particle relaxation time against Mach strength}
    \label{fig:machvsRe}
\end{figure}

Three experimental conditions were targeted, based on previous simulation work \cite{mcfarland2016computational}, in which particles will have small fast reacting particles with very low initial momentum deficit ($\tau_{v,c} < 0.1$), medium transitional particles with small momentum deficit ($1<\tau_{v,c}<10$), and large particles with high initial momentum deficit ($\tau_{v,c}>10$). Given the particle materials and sizes available from suppliers these times had to be modified. The best fit for the fast relaxation time was titanium dioxide ($\rho_{Ti}=4200 [kg/m^3]$ \cite{USNANO}) with a mean particle diameter 0.2 [$\mu m$]. For the medium and low relaxation times aluminum ($\rho_{AL} = 2700 [kg/m^3]$) with mean particle diameters of 1.56 [$\mu m$] and  2.31 [$\mu m$], see table \ref{table:IC_At_table} and figure \ref{fig:machvsRe}.

\subsubsection{Particle size characterization} \label{sec:solid_part_char}

Scanning electron microscopy (SEM) was used to confirm the initial particle shapes and sizes, as prepared. SEM imaging was crucial for accurately characterizing the particles under investigation. The morphology of the particle becomes more important as the particle's size increases, ultimately modifying the drag coefficient and its ability to follow the flow. Due to this, a series of SEM images were acquired in the Microscopy Imaging Center at Texas A\&M University to investigate the particle size and shape distribution. The general procedure involved taking a sample of the particles, dispersing them in an alcohol solution, and sonicating them at 40 [kHz] to reduce agglomeration. Afterward, a drop of the suspension was deposited onto a carbon adhesive tape, mounted on the SEM stub, and dried at room temperature. Samples were then sputter coated with 10 [nm] gold, using Cressington 108 sputter coater, to render the conductivity of the samples and minimize charging. A sequence of images at multiple locations within the sample was obtained to build a statistical representation. 

\begin{figure}[ht]
	\centering
		\includegraphics[width=1\textwidth]{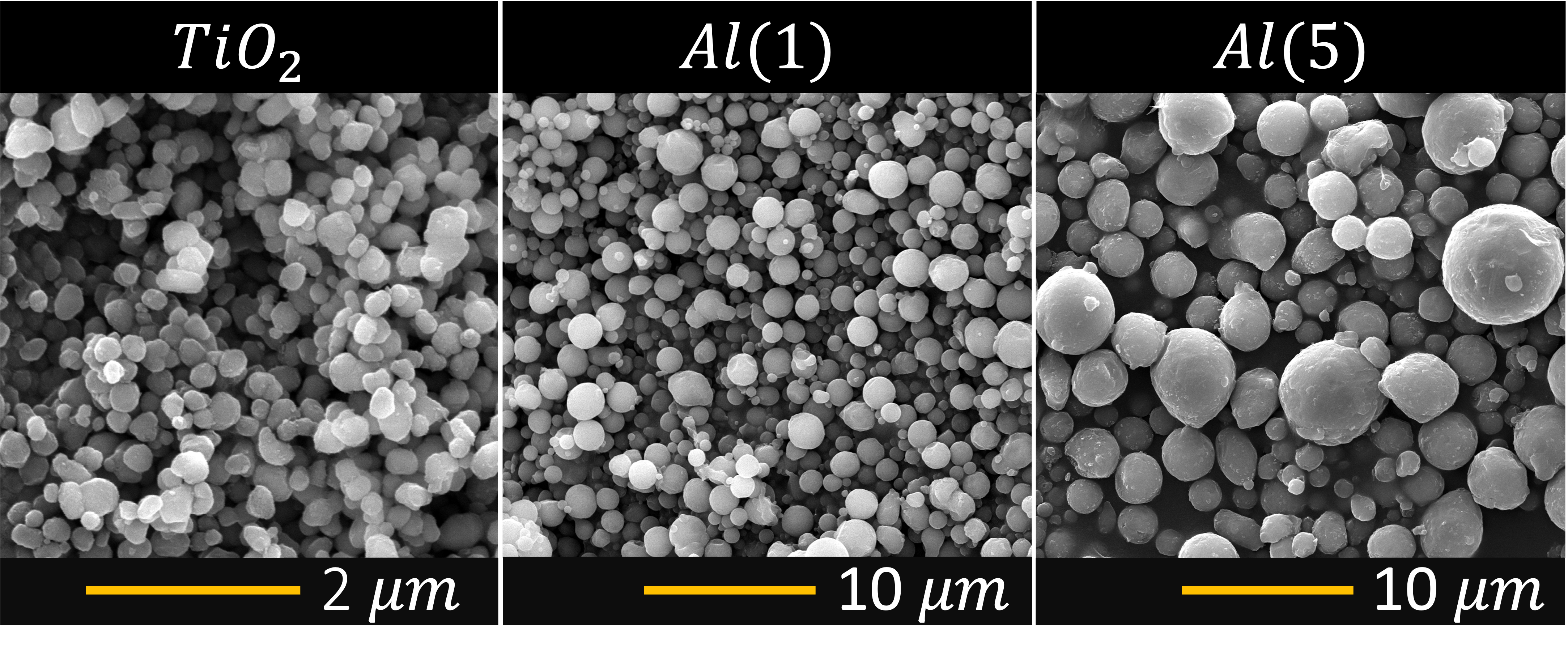}
	  \caption{SEM imaging. Note: The left figure is on a different scale.}
	  \label{figure:SEM}
\end{figure}

Imaging was done using the FEI Quanta 600 SEM, a 30 kV accelerating voltage, and the Secondary Electron (SE) detector. Higher magnifications were required depending on the size of the particle. Figure \ref{figure:SEM} shows a sample of the SEM image results collected for the three selected particles. The images where pre-processed and analyzed in MATLAB utilizing a circle finder algorithm detecting the edge of the individual particles. From the collected results, we can conclude that the particles can be considered effectively spherical with their respective distribution, as shown in figure \ref{figure:SEM}. 

It is important to emphasize that micron-sized particles tend to clump or agglomerate into clusters that act and are measured as effectively larger particles dispersed into the interface. In addition to scanning electron microscopy (SEM), the seeded particle distributions were thoroughly validated using a TSI inc. Phase Doppler Particle Analyzer (PDPA) to measure particle sizes in-situ. The PDPA was set to detect reflection as the predominant scattered mode for the solid particles. Multiple data collections were performed before conducting the experiments. Particles were measured \textit{in-situ} utilizing the experimental procedures as will be described in the next section (\S \ref{sec:Init_cond}). This step ensured a comprehensive understanding of the initial conditions. For more detailed information on the setup and configuration of the PDPA, please refer to \cite{duke2023experiments}.


The closest representation of the initial particle shape and size can be obtained from the SEM results. However, these conditions are not entirely representative of the true experimental conditions. Generally, SEM results indicate particle sizes smaller than those recorded with PDPA. This discrepancy can be attributed to the efficiency of the ejector pump in separating particle clumps. As can be seen in table \ref{table:Particle size staticstics}, the SEM confirmed that $TiO_2$ had a mean particle size on the order of 0.2 $\mu m$, which makes them ideal passive flow tracers at our conditions. Unfortunately due to their size, they could not be detected with PDPA since the minimum detectable size is 0.5 $\mu m$. On the other hand, PDPA measurements of the aluminum cases, Al(1) and Al(5), indicated higher particle sizes compared to the SEM due to agglomeration. We take the PDPA measurements to provide the most accurate representation of the particle sizes for theoretical analysis of the SDMI. Thus, all referenced particle diameters (e.g. $d_{10}$, $d_{32}$) will be from the PDPA data. 

The particle size uncertainty depends on the optical technique utilized. In the case of the SEM imagery,  uncertainty for the particle distribution was at minimum 2 pixels, corresponding to 3.9  $[nm]$, 19.5  $[nm]$, and 58.6 $[nm]$ for the $TiO_2$, Al(1) and Al(5) particles, respectively. PDPA accuracy is constrained by the largest particle size that the lens and the sampling volume can detect. The 135 [mm] focal length lens used can resolve 0.5-40 $[\mu m]$. The uncertainty depends on the largest measurable diameter ($D_{x}$) and the diameter that was actually measured ($D_m$), calculated as $\pm 0.01(D_{x} + D_m)$. In other words, the minimum particle diameter (0.5 $[\mu m]$) that can be measured with the 120 mm lens ($D_x=44.83 \mu m$) will have an uncertainty of $\pm 0.4533 [\mu m]$.

\begin{table}[h!]
\centering
\caption{Particle size statistics}
\begin{tabular}{|cc|c|c|c|}
\hline
\multicolumn{2}{|c|}{PI}                               & $TiO_2$ & $Al(1) $ & $Al(5) $ \\ \hline
\multicolumn{1}{|c|}{{SEM}}  & $D_{10}[\mu m] $ & 0.2     & 1.18      & 4.61      \\ \cline{2-5} 
\multicolumn{1}{|c|}{}                      & $D_{32}[\mu m] $ & 0.3     & 1.65      & 6.28      \\ \hline
\multicolumn{1}{|c|}{{PDPA 120 mm}} & $D_{10}[\mu m]$ & -       & 1.56      & 4.304      \\ \cline{2-5} 
\multicolumn{1}{|c|}{}                      & $D_{32}[\mu m]$ & -       & 2.31       & 9.73     \\ \hline
\end{tabular}
\label{table:Particle size staticstics}
\end{table}

\subsubsection{Initial conditions} \label{sec:Init_cond}

\begin{figure*}[ht]
	\centering
		\includegraphics[width=1\textwidth]{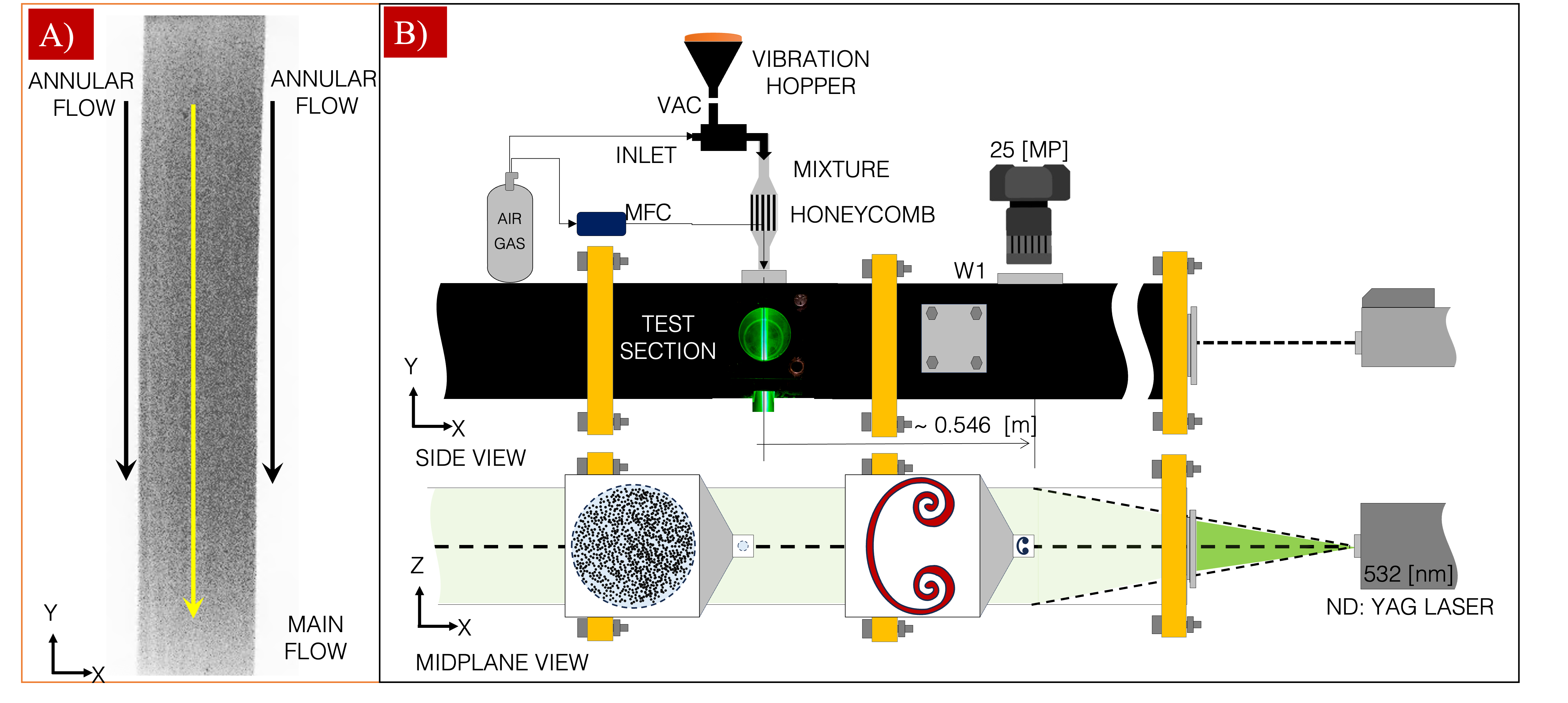}
	  \caption{Interface Initial Conditions and Diagnostics: A) example of the stability of the multiphase interface, and B) test section configuration}
	  \label{figure:IC_laser}
\end{figure*}

A thorough understanding of the initial conditions is necessary for validation of SDMI theory. This requires measurement of the multiphase interface properties and the particle and gas properties in-situ. The stability of the multiphase interface was analyzed using a time sequence Mie-scattering imaging of the X-Y plane, shown in figure \ref{figure:IC_laser}. Various gas flow rates were experimented with to determine the stable laminar regime. The ejector pump was set at 6 psi, yielding a total flow rate of 5.75 [SLM], and the annular flow was set to 4 [SLM], conditions found to produce a stable interface. A set of interface X-Y plane images were captured and post-processed to quantify the interface dimensions. These measurements showed the width of the interface to be for the three cases  $D_{TiO_2} \sim 12.37 [mm]$, $D_{Al(1)} \sim 12.49 [mm]$  and $D_{Al(5)} \sim 12.37 [mm]$, with an average width of the interface of $ D \sim 12.41 \pm 0.11 [mm]$.

\begin{table}[h!]
\centering
\caption{Experimental Post-Shock Conditions}
\begin{tabular}{|c|c|c|c|} 
\hline
Mach & V [m/s] & P[kPa]& T [K]\\
\hline
$1.35 \pm 0.007$& $177.4 \pm 3.3$& $196.1\pm 2.2 $& $359.73 \pm 1.32$\\ 
\hline 
\end{tabular}
\label{table:MACH,P,T,V}
\end{table}


The experiments were conducted with three different $\tau_v $, with $A_e$ held constant. The reference case, the small fast-reacting particle case, utilized  $TiO_2$ particles with $\tau_{v,c} < 0.1 $. For this short relaxation time, particle lag effects could not be distinguished. The particle number density required to reach $A_e \sim 0.04 $ in this case were prohibitively high, resulting in a dense particle field that prohibited the measurement of velocities. Thus, we chose to use a gas density difference ($A_g \sim 0.04 $) to provide the required $A_e$ while the $TiO_2$ particles acted as passive tracers, essentially creating a pure RMI. The carrier gas was a mixture $N_2$ and $CO_2$ (80/20 by volume). For the aluminum Al(1) and Al(5) cases, $A_e$ was entirely controlled by the particle mass concentration, and both the carrier and surrounding annular gas were air ($A_g \sim 0 $).

\begin{figure}[ht]
	\centering
		\includegraphics[width=0.8\textwidth]{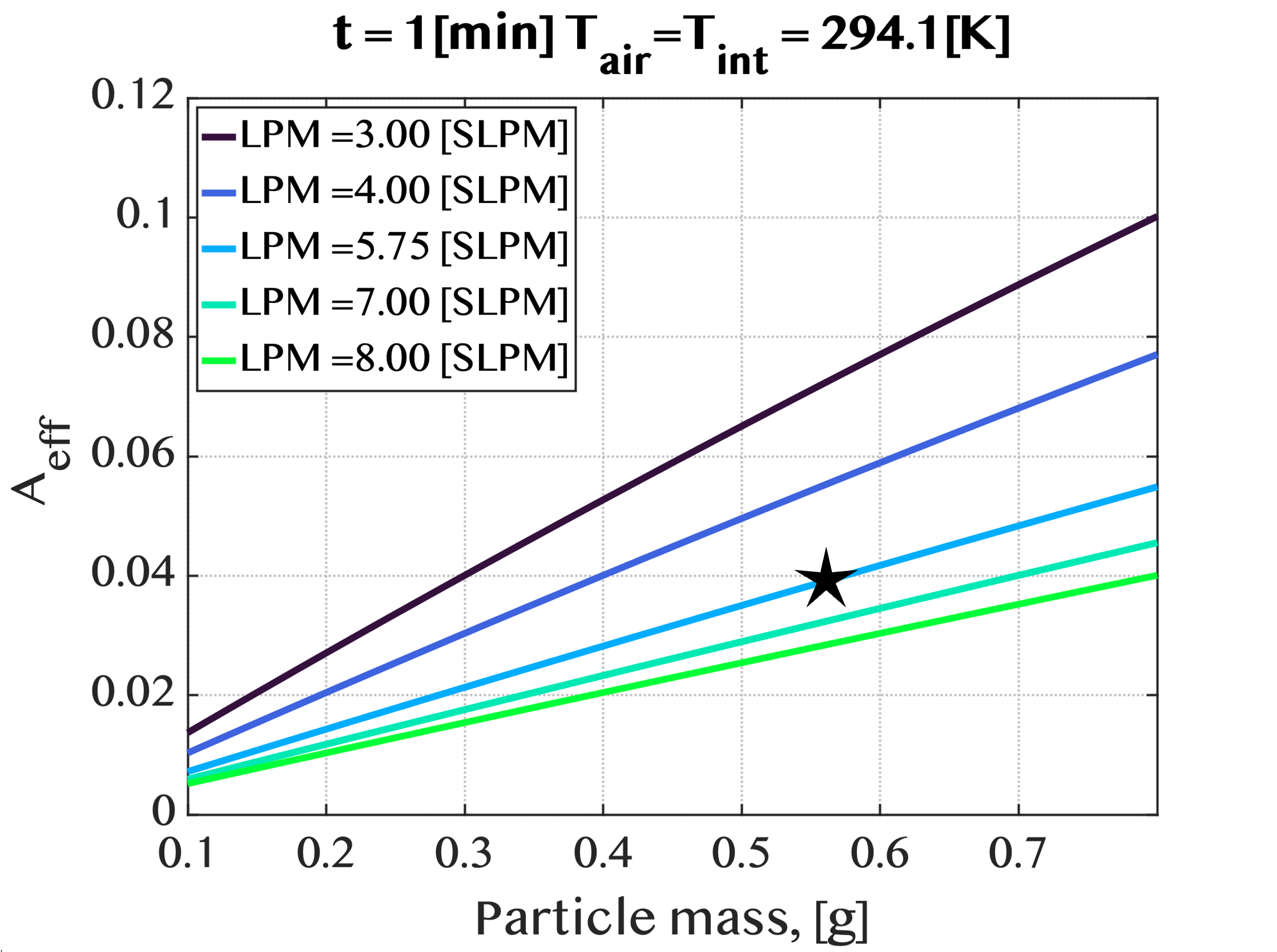}
	  \caption{Experimental parameters for the effective Atwood number $A_{eff}$}
	  \label{figure:A_eff}
\end{figure}

Particle concentration at the initial condition, as shown in figure \ref{figure:A_eff}, relies on two key parameters: the gas flow rate and the particle discharge rate. In the Al(1) and Al(2) cases, mass flow from the hopper was held at a nearly constant output, and the gas flow rate varied. By balancing these parameters, the effective Atwood number can be prescribed. As shown in table \ref{tab:Hopper_Settings}, several small changes in the vibrator eccentric mass and orifice diameter were required to achieve near-constant vibration hopper mass output. Once this is guaranteed, the mass is transported with the ejector pump into the flow straightening section, creating a stable interface. Then, the particle concentration of the interface was studied utilizing a mass filtration system, described in \cite{duke2020method}. For the $TiO_2$ case, much lower particle concentrations were required. Thus, a commercial fluidized-bed PIV seeding device (TSI inc. model 9309) was used for the interface gas. The vibration hopper was used to seed the annular co-flow gas for diagnostic purposes (PIV, see \S \ref{sec:exp_results}) by operating it in a pulsed mode, activating it for 1 s then allowing the resulting particles to flow into the shock tube for the next 5-10 [s], until the experiment is run. 

The particle mass was retained with a filter pore size of 0.1 $[\mu m]$. Samples were collected for $\sim 1 [min]$ and weighed with an analytical balance with an accuracy of $\pm 0.1 [mg]$ to estimate the average particle concentration in the interface mass. Each case was measured fifteen times for greater statistical certainty. The $TiO_2$ case was found to have negligible mass, as desired, in the interface and annular gas flows. As shown in section \ref{sec:Exp_sens_analysis},  even the smallest variability in temperature alone could affect $A_e$ through variations in the gas density. For this reason, two thermocouples with an accuracy of  $\pm 1 [C]$ were used to record the surrounding air $T_1 = T_{surr}$ and the carrier gas $T_{int}$ temperature just before the initiation of the shock wave. Utilizing the gas volume and density, and the particle mass and volume displaced, the effective Atwood number and the uncertainty were calculated.

\begin{table}[]
\centering
\begin{tabular}{|l|c|c|c|}
\hline
System Configuration          & $TiO_2$ & $Al (1) $ & $Al (5)$ \\ \hline
Orifice Diameter [mm]& 2.0& 1.5       & 1.0\\ \hline
Eccentric mass [g]& 3.47    & 4.93      & 3.47      \\ \hline
Discharge rate [g/min]& 0.751& 0.837& 0.875 \\
\hline
Operation time [s] & 1 & continuous & continuous 
\\ \hline
\end{tabular}
\caption{Hooper settings}
\label{tab:Hopper_Settings}
\end{table}



\begin{figure}[ht]
	\centering
		\includegraphics[width=0.8\textwidth]{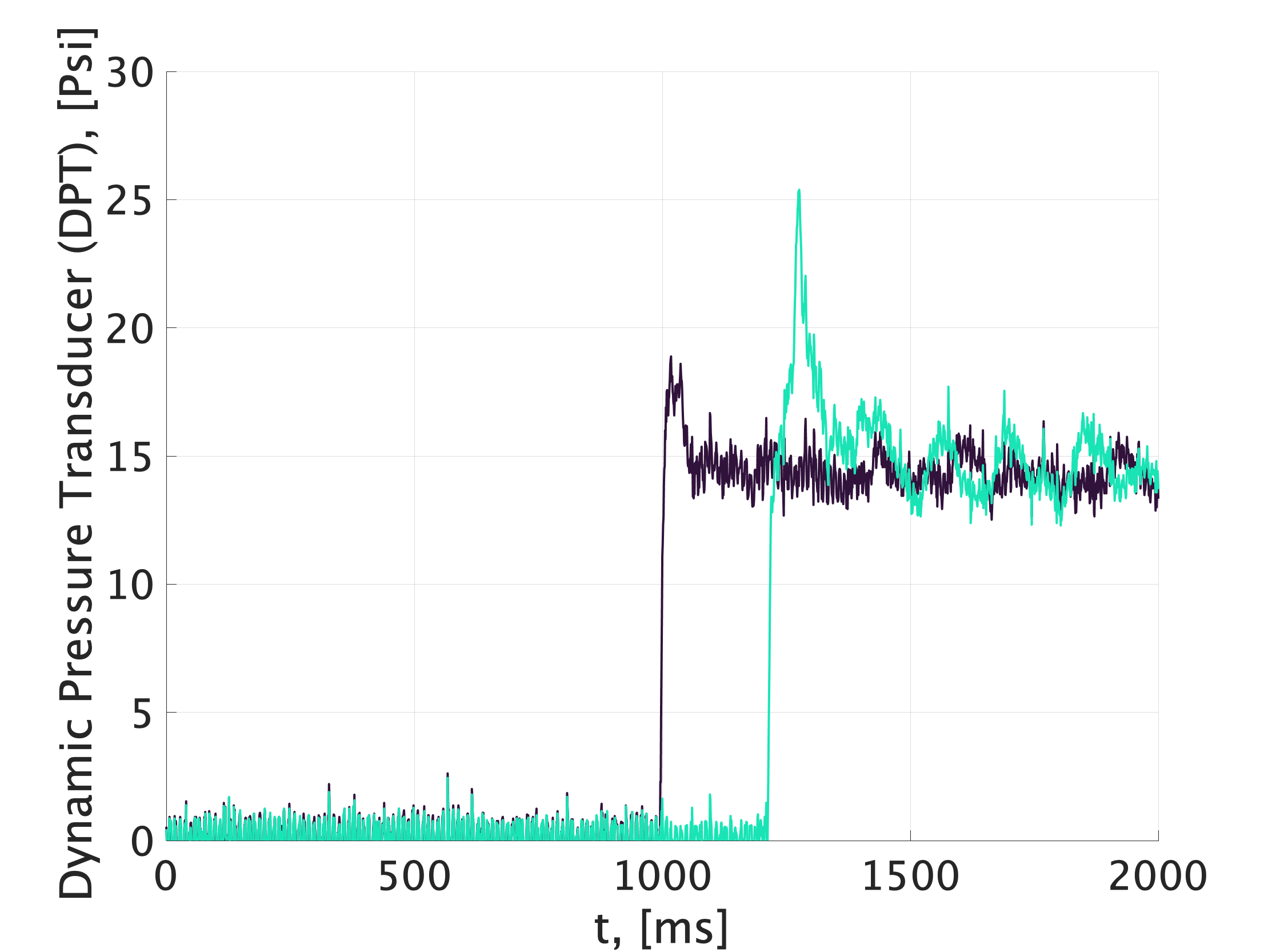}
	  \caption{Experimental pressure signals}
	  \label{figure:DPT_plot}
\end{figure}

The post-shock conditions were then estimated by gas dynamics equations from the initial gas temperature of the driven section $T_1$ and the shock wave velocity $w_i = \frac{\delta_x}{\delta_t}$. The shock wave velocity was found from the time-of-arrival $\delta_t$ across two piezoelectric transducers spaced a distance of $\delta_x$ apart. The arrival time is marked by when the pressure registers at 5 [\%] of the pressure peak value. Pressures are measured at a rate of 1.25 MHz, resulting in a maximum error of $\pm 800 ns$. An example of the experimental pressure measurements can be seen in figure \ref{figure:DPT_plot}, where the shock arrival is visible as the rapid rise in pressure. The pressure signals are also seen to exhibit some overshoot and ringing, but these do not affect the accuracy of the arrival time measurement. A summary of the initial conditions for each case can be found in table \ref{table:IC_At_table}.



\begin{table}
\begin{ruledtabular}
\caption{Experimental Initial Conditions. Note: Stokes/Cloutman (S/C) nondenominational relaxation time.}
\begin{tabular}{lcdr}
\textrm{PI}&
\textrm{$TiO_2$}&
\multicolumn{1}{c}{\textrm{$Al (1)$}}&
\textrm{Al (5)}\\
\colrule

$\tau_v$ (S/C) for $d_{10}$ & 0 & 3.01/2.22 &  6.61/4.56\\

$\tau_v$ (S/C) for $d_{32}$  & 0 & 22.94/13.99 & 117.26/58.27\\

$T_{int}$ & 19.42& 20.37 & 19.78\\

$T_{surr}$  & 19.66& 19.93& 19.81\\

$m_{p}$  & $\sim 0$& 0.542& 0.534\\

$A_g$ & 0.0388 & 0 & 0\\

$A_{e}$ & 0.0388 & 0.0376 & 0.0371\\

$\sigma \pm$ & 0.0001 & 0.0086 & 0.0074\\
\end{tabular}
\label{table:IC_At_table}
\end{ruledtabular}
\end{table}

\subsubsection{Imaging methods}  \label{sec:imaging_methods}

Several steps were taken to ensure consistency between experiments. Alignment of the laser, optics, and cameras were all checked with calibration targets, more detail on \cite{duke2020method}. Optical equipment and windows are cleaned to remove any collected particle residues or dust. Lasers and cameras were focused and aligned to an X-Z plane with sufficient distance from the walls to avoid boundary effects. The AL(1) and AL(5) cases were imaged at the midplane, while the $TiO_2$ case was imaged at $\sim 1 cm$ from the top boundary, limiting the effect of gas diffusion (uses $N_2/CO_2$ for carrier gas) on the interface. Interface gas diffusion increases with jet (the cylindrical interface) development length, while the particles are relatively free of diffusion.

\begin{figure}[ht]
	\centering
		\includegraphics[width=0.8\textwidth]{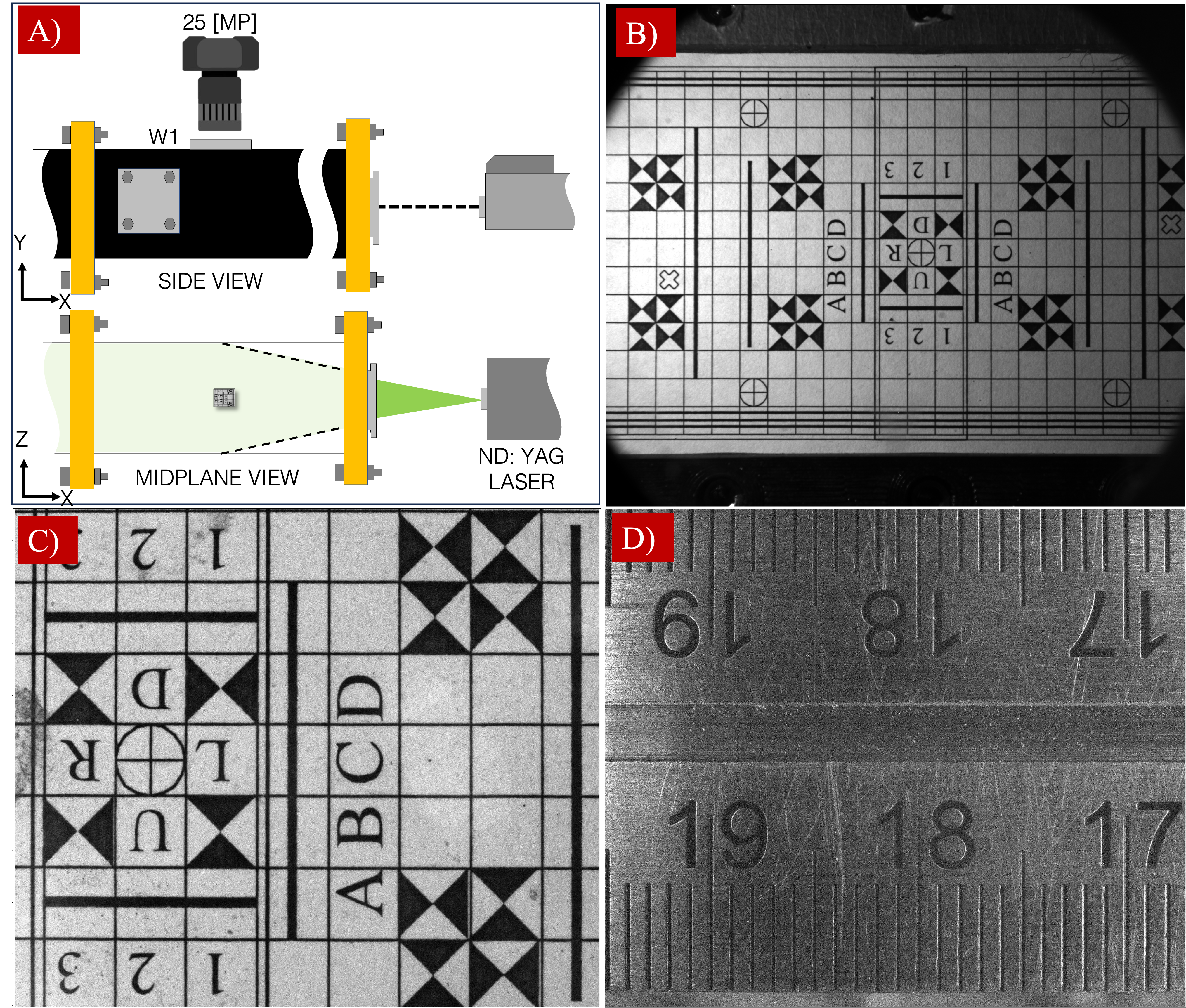}
	  \caption{Camera Settings for the experimental setup: A) Top camera setting, B) Wide FOV, C) Higher Magnification small FOV image, and D) calibration image}
	  \label{figure:CAMERAS POSITION} 
\end{figure}

Cameras were focused first at a f/stop of 1.8, the mid-range of the hyper-focal distance, and then adjusted using particle Mie-scattering signal from the laser.  The timing and location of the developing multiphase interface were established with a wide FOV (MAG = 0.1 ) camera as seen in figure \ref{figure:CAMERAS POSITION}. With timing established, higher magnifications (MAG = 0.5) were used to provide better discrimination of scattered light from individual particles, reducing 3D effects and improving contrast for the PIV method. Depending on the experimental position, cameras must be adjusted to avoid overexposure or over-saturation of the camera sensor. Calibration images were captured for each camera setup, obtaining a scale to transform from [pixel] to [mm] and providing an image position relative to the shock tube, see table \ref{table:camerasettingstable}. They also provided an image map for correcting any optical distortions. Camera settings are shown in figure \ref{figure:CAMERAS POSITION}. Background images are also taken before each experiment to account for any changes in ambient light in the experimental setup.

\begin{table}
\centering
\caption{Camera settings}
\begin{tabular}{|c|c|c|c|}
\hline
Cases   & MAG  & FOV ([mm]x[mm])  & [$\mu m$/px] \\   \hline
$TiO_2$ & 0.4  & 36.70x31.76      & 6.93          \\    \hline
Al(1)   & 0.47 & 31.24x27.04      & 5.90          \\   \hline
Al(5)   & 0.29 & 50.62x43.82      & 9.56          \\   \hline
\end{tabular}
\label{table:camerasettingstable}
\end{table}

\subsubsection{Experimental procedure and data acquisition}
\label{sec:exp_data_proc}


Before running a new experiment, the processed particles are set into the vibration hopper to be dispersed and used within one to two hours to prevent atmospheric moisture intrusion. The diaphragm is replaced and the driven section is vacuumed to reduce any debris or spurious particles from the previous experiment. The hopper and gas mass flow rates are then started and allowed to reach a steady state operation (after two minutes as previously described) before the experiment is run. For the $TiO_2$ case (gas Atwood number), the fluidized-bed PIV seeding device is charged first with particles to seed the interface gas and the gas flows started. For experiments where PIV images are acquired, the vibration hopper is activated for 1s to charge the annular gas flow with tracer particles. 

An experiment is then initiated utilizing our LabVIEW Automatic Shock Firing Sequence (ASFS) to ensure consistency and precision. The sequence is started once the interface gas flows have been initiated. The ASFS then slowly fills the driver pressure through a small solenoid valve until it reaches the target static pressure. Once this pressure is reached, a large solenoid valve is activated to rapidly raise the pressure (within $\sim 0.5 [s]$) and rupture the diaphragm (see \S \ref{sec:Experimental Facility}). Once the diaphragm breaks the shock is created and detected by the first dynamic pressure transducer, triggering the two laser pulses and camera frames (A \& B) to acquire image pairs for PIV. These steps must are repeated for 10 experimental trials for each case,  to ensure repeatable morphological development of the interface. For more details, see \cite{duke2023experiments,duke2020method}. 

PIV images were analyzed utilizing DAVIS 11.0 software \cite{LaVision_Inc} with an inter-frame time of $\Delta t = 2 [\mu s]$, ensuring accurate correlations at the experimental velocities. A multi-pass cross-correlation of the PIV images was used to analyze the movement of the particles. The cross-correlation between images was initially performed using interrogation regions of 512 x 512 pixels. In the final four passes, the size of the interrogation region was reduced to 96 x 96 pixels. In both passes, the overlap of interrogation regions was set to be 50\% and 75\%. After calculating the velocity vectors, they were post-processed with a signal-to-noise medium threshold. A normalized peak height filter of 1.5 was employed to identify other outlier vectors. After calculating the velocity vectors, they were post-processed with a 5x5 median filter, disregarding the outlier vectors, and replacing them with interpolating vectors. The overall result was a grid size of 24 x 24 pixels for each case. Afterward, the mean of the velocity vector in both directions is then subtracted from the velocity field to obtain relative velocity vectors to be analyzed for vorticity estimation. Finally, an 11x11 smoothing Gaussian filter is applied to the velocity field to remove spurious noise. 

\section{Experimental results}
\label{sec:exp_results}

This section will detail the results, both qualitatively and quantitatively, of the development of the multiphase interface. A description of the interface evolution and comparison of the multiphase development is provided from the imagery. From PIV results, an estimate of the circulation deposition is provided and compared with RMI theory. 


The hydrodynamic development starts with a sudden compression of the cylindrical interface that then evolves after the interface is accelerated, forming two counter-rotating vortices. The shock wave will cause the gas to jump instantaneously into post-shock conditions, while the particles are initially unaffected. The smaller particles achieve equilibrium rapidly, whereas larger particles respond slower, lagging behind and stratifying by size. As the gas accelerates the particles, they start transferring momentum and energy with the surrounding gas.

The gas experiences a velocity jump of approximately 177.4 [m/s] (derived from the piston velocity in 1D gas dynamics and post-shock conditions detailed in section 3.3). However, as seen in table \ref{table:PIV_DATA}, the particle velocities are lower in all cases. Cases 1 and 3 showed a larger velocity deficit from the average gas velocity ($<10\%$) at late times ($\sim 3 [ms] $). Some amount of velocity deficit should be expected as the higher effective density of the interface gas-particle mixture requires more momentum to accelerate. This deficit should be reduced with time as the interface gas and surrounding equilibrium. For case 1, the interface gas will jump to a slightly lower velocity, due to the gas mixture properties and the high level of mixing. For case three, the equilibration time is on the order of milliseconds for the largest particles present, and the interface may not have obtained equilibrium. 

\begin{table}
\centering
\caption{Mean particle velocity and shock strength estimation from PIV results and pressure transducers}
\begin{tabular}{|c|c|c|c|}
\hline
Case     & $TiO_2$           & $Al (1) $         & $Al (5) $         \\ \hline
M        & $1.347 \pm 0.0062$ & $1.354 \pm 0.004$ & $1.349 \pm 0.008$ \\ \hline
$V_{gx}$ & $172.998\pm 2.74$  & $176.19\pm 1.91$  & $173.80\pm 3.56$  \\ \hline
$V_{px}$ & $164.32 \pm 1.24$ & $175.01 \pm 1.41$ & $166.63 \pm 1.21$ \\ \hline
$V_{py}$ & $-0.87 \pm 0.36$  & $-1.49 \pm 1.30$  & $-0.81 \pm 0.15$  \\ \hline
\end{tabular}
\label{table:PIV_DATA}
\end{table}

\begin{figure*}[ht]
	\centering
		\includegraphics[width=1\textwidth]{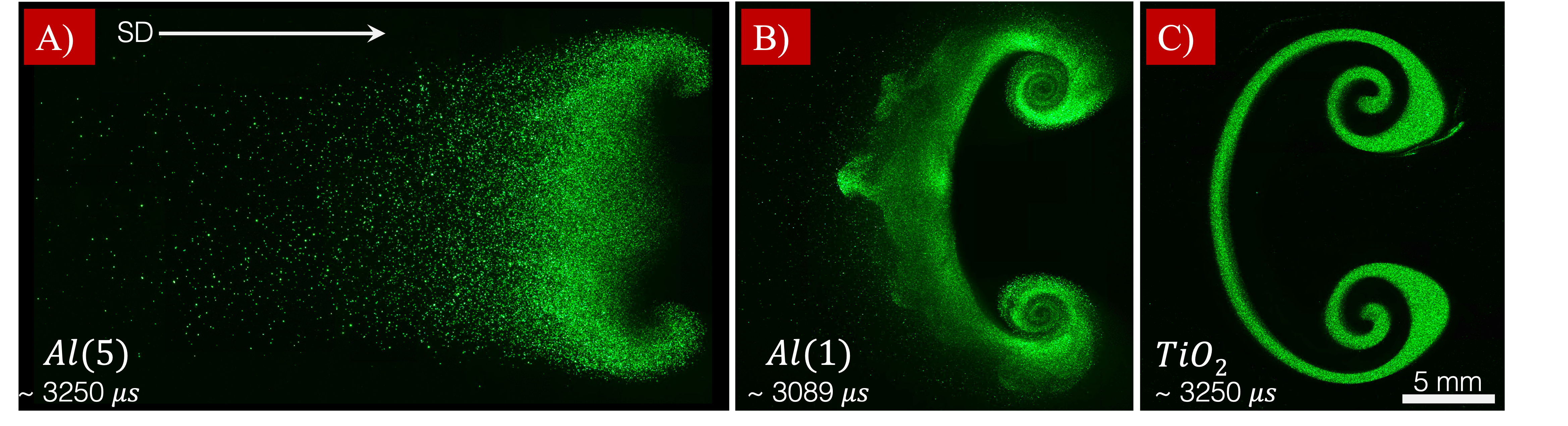}
	  \caption{Interface Evolution for the Shock-Driven Multiphase Instability for the A) large particle, B) small particles, and C) gas cases.}
	  \label{figure:PL_Evolution_exp}
\end{figure*}

As can be seen in figure \ref{figure:PL_Evolution_exp} the results are highly symmetric for the classical RMI case ($\tau_v \sim 0$). The small $TiO_2$ particles appear to exchange momentum and couple with the gas without interference or modification to the flow. Qualitatively comparing this case with the Al(1) and  Al(5) cases, the effects of the particle response time on the overall hydrodynamics can be seen. As the non-dimensional particle response time increases, the interface development decreases and the length of trailing particles increases. 

As shown in figure \ref{figure:PL_Evolution_exp} for the Al(1) case,  the overall interface shows a similar morphology to the RMI case; however, the vortex cores are less well defined as particles stratify (spread out) due to the centripetal acceleration. In addition, the upstream interface shows some additional perturbations and greater thickness due to particle lag effects. Lastly, in the Al(5) case, very little development is observed, followed by a long region of trailing particles, referred to as the particle tail \cite{middlebrooks2018droplet}. This region is created as larger particles fall the furthest behind, effectively stratifying by size, and is more pronounced as the average particle size and standard deviation increase.  Noticeably, the particle's long relaxation time strongly dampens the hydrodynamics mixing of the interface.

\begin{figure*}[ht]
	\centering
		\includegraphics[width=1\textwidth]{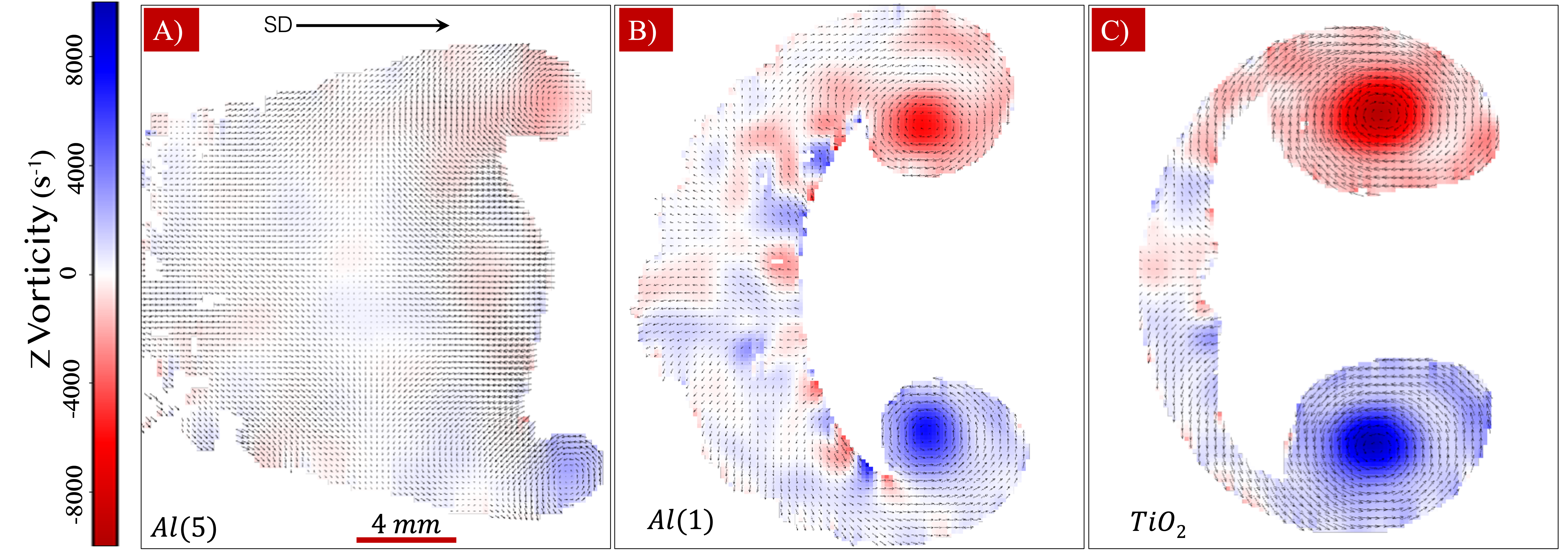}
	  \caption{Vorticity Fields for the A) large particles, B) small particles, and C) gas cases.}
	  \label{figure:PIV_vorticity}
\end{figure*}

Velocity fields are acquired by PIV as described in \S \ref{sec:exp_data_proc} with the mean velocities shown in table \ref{table:PIV_DATA}. For the Al(1) and Al(5) cases, the particles contained in the interface gas were found to spread sufficiently to the adjacent surrounding gas as to provide reliable PIV vectors. For the $TiO_2$ case, particles had to be added to the annular co-flow gas to provide a sufficient number of velocity vectors. As the $TiO_2$ particles were in very low concentrations, measured to have negligible mass fraction, they did not alter the evolution of the interface.  

From the velocity fields, the vorticity ($\omega_{z}$) is found through a first-order finite difference approximation , and is presented in figure \ref{figure:PIV_vorticity}. The vorticity fields are plotted on the same scale, but it should be noted that the $TiO_2$ case vorticity peaks exceed the bounds of this scale. Even considering this, it is clear that the RMI case shows much stronger vorticity than the other cases. Apart from the intensity, the size of the vortex cores also decreases as the particle relaxation time increases. This indicates a lower mixing rate due to the delayed particle response. 

To numerically estimate this damping of the mixing rate, the circulation was estimated from the vorticity fields for each case. Several methods were used to analyze the location of the vortex core and variations in the location of the vortex core. The first method estimated the vortex core location from the maximum intensity of the swirling strength. The second method is borrowed from theory \cite{archer2008numerical} when considering well-formed vortex rings. The vortex ring is parameterized by \cite{ames2023multifluid} on their size and swirling strength. Assuming axisymmetry, the measurement of the ring radius can be derived from the first and second radial moments of the azimuthal vorticity eqns \ref{eqn:Azi1} and \ref{eqn:Azi2}. The calculation of the vortex parameters \ref{eqn:Azi1}, \ref{eqn:Azi2} and \ref{eqn:zc} are obtained 
for each core from the vorticity fields which are estimated from the local vorticity to the pixel area $dA = \Delta x \Delta z$, and then averaged over the entire circumference area of the maximum vorticity. 

\begin{equation}
    R = \frac{1}{\Gamma}\sum_{i,j}^{} \omega_{i,j} r_idA 
\label{eqn:Azi1}
\end{equation}

\begin{equation}
    R_{2}^{2} = \frac{1}{\Gamma}\sum_{i,j}^{} \omega_{i,j} r_i^2 dA 
 \label{eqn:Azi2}
\end{equation}

\begin{equation}
    z_c = \frac{1}{\Gamma}\sum_{i,j}^{} \omega_{i,j} z_j dA 
 \label{eqn:zc}
\end{equation}

\begin{figure}[ht]
	\centering
		\includegraphics[width=0.8\textwidth]{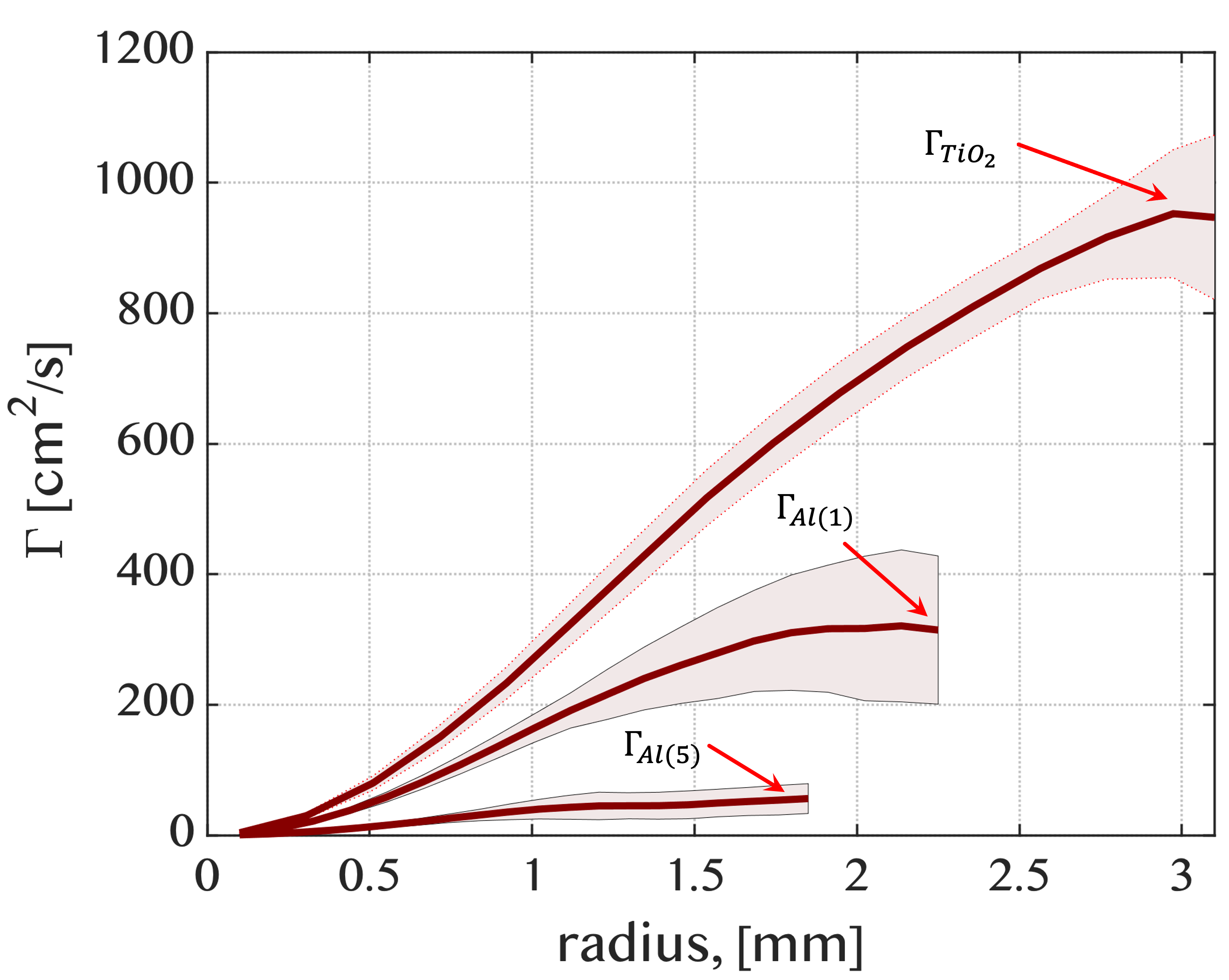}
	  \caption{Circulation estimation from the center of the vortex core}
	  \label{figure:PL_cir_exp_radius}
\end{figure}

Once the vortex core location is found, the circulation is obtained by tracing twenty concentric closed-path circles around it. By definition, the circulation can be obtained by utilizing the tangential velocity vector component, see eqn \ref{eqn:Gamma_tang}) in which N is the number of divisions of the circle. If we take constant length ($l_i = 2 \pi r/N$), the mean circulation is obtained for each closed path around the center of the vortex. The maximum circulation is obtained for each case from the captured velocity vectors at each circle radius, as seen in figure \ref{figure:PL_cir_exp_radius}.

\begin{equation}
   \Gamma =  \oint\ \vec{V_{\theta}}  \cdot \vec{ds} \approx \sum_i^N V_{\theta, i} l_i
\label{eqn:Gamma_tang}
\end{equation}

The average of the maximum circulation obtained for each trial is used to estimate the circulation for each case. These values are shown in table \ref{table:Circulation} along with the standard deviation of the trials. 

\begin{table}[h!]
\centering
\caption{ Circulation estimation from PIV results}
\begin{tabular}{|c|c|c|c|c|} 
\hline
$\Gamma$ & $TiO_2$ & $Al (1) $& $Al (5) $\\
\hline
{$\Gamma_{mean}$} & 952.45 & 320.75 & 56.159\\
\hline
$\sigma \pm$ & 98.078 & 116.46 & 22.85\\
\hline
\end{tabular}
\label{table:Circulation}
\end{table}

\section{Modeling}
\label{sec:modeling} 

This section describes how the particle lag effects were modeled for the experimental conditions to predict the circulation. The effect of particle size (lag distance) on circulation is derived from the vorticity (eqn \ref{eqn:vort}) and enstropy (eqn \ref{eqn:enst}) equations first. Then this model is fitted to the circulation predicted for an RMI (eqn \ref{eqn:PBcirc}) in the limiting case of infinitely small particles. 

\begin{figure}
\includegraphics[width=0.6\textwidth]{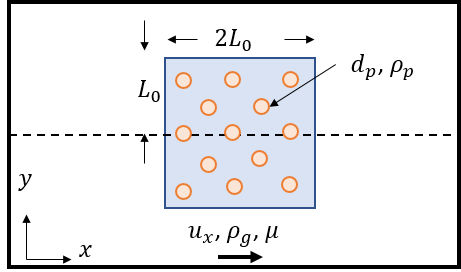}
\caption{\label{fig:cv} Example control volume for vorticity generation by particle source term.}
\end{figure}

Equation \ref{eqn:vort} can be simplified by taking a control volume that travels with the particles. For simplicity, we take the control volume to be a square region with sides of length $2L_0$ containing uniform particles with a total mass of $m_p$, see fig. \ref{fig:cv}. The flow can be assumed to be 2D and incompressible after the initial passage of the shock wave. For a pure SDMI, the baroclinic term can be neglected as there are no gradients in the gas density. The velocity, $u_i$, can be taken as the relative velocity of the gas to the particles. This can be simplified by assuming that the velocity and momentum source terms are predominantly in the x-direction, neglecting the y-components. With these assumptions, the vorticity local to the control volume, ($\omega_l$) follows eqn. \ref{eqn:vort_simp}. Likewise the enstrophy equation can be reduced to find the total enstrophy generated in the control volume, including that which is advected away, as shown in eqn. \ref{eqn:enst_simp}.

\begin{equation}
\label{eqn:vort_simp}
  \partial_t \omega_l = \frac{1}{\rho_B'}\partial_y B_x -u_x \partial_x \omega_l 
\end{equation}

\begin{equation}
\label{eqn:enst_simp}
  D_t \Omega = \frac{1}{\rho_B'} \omega_l \partial_y B_x
\end{equation}

To determine the total enstrophy generated, it is necessary to find a solution for $\omega_l(t)$, which requires additional simplifications. Since vorticity production is antisymmetrical (positive/negative cores) the control volume is taken as one-half the interface with volume of $2L_0^2 dz$, where $dz$ is taken as a unit depth. The divergence theorem is applied to the advection term and integrated over $z$ (constant), resulting in a line integral over the y-direction. Since there is no vorticity upstream, only the negative flux along the downstream boundary remains. The source term is then integrated over $z$ (constant), and Green's theorem applied resulting in a line integral of $B_x$ along the lower boundary. Evaluating the integrals and rearranging, equation \ref{eqn:vort_flux} is obtained, in which $K_1=1/(2L_0)$ and $K_2=1/(L_0 \rho_B')$.

\begin{equation}
\label{eqn:vort_flux}
  \partial_t \omega_l + K_1 u_x \omega_l = K_2 B_x 
\end{equation}

Eqn. \ref{eqn:vort_flux} is in the form of an inhomogeneous first-order linear ordinary differential equation (ODE). To find a solution for $\omega_l(t)$, the method of variation of parameters is applied. The solution for the homogeneous ODE is found first as $\omega_l(t)=A e^{-K_1 x(t)}$. The solution to the inhomogeneous ODE may then be found in the form of $\omega_l(t)=q(t)e^{-K_1 x(t)} + A e^{-K_1 x(t)}$, where $q'(t)= K_2 B_x(t) e^{K_1 x(t)}$. To solve this equation a function for $B_x(t)$ is needed. Utilizing the Kliatchko drag model (eqn \ref{eqn:drag}) results in an equation ($q'(t)= K_2 m_p B (x'(t) + C x'(t)^{5/3}) e^{K_1 x(t)}$) for which there is no closed form solution for $q(t)$. 

Instead, the Stokes drag formula ($B_x=m_p B x'(t)$) is used. Applying this equation to the inhomogeneous ODE we get $q'(t)=K_2 m_p B x'(t) e^{K_1 x(t)}$, which has a solution of $q(t)=K_2 m_p B e^{-K_1 x(t)}/K_1$. A solution for the inhomogeneous ODE may be found as shown in eqn. \ref{eqn:vort_loc}. The equation presents itself in the form of the ag distance $x(t)$. From the Stokes drag law, the lag distance can be found as $x(t)=-(u_0/B)(e^{-Bt}-1)$. The maximum vorticity is obtained as $t \to \infty$, when the lag distance reaches its maximum value, $x_s$ (eqn \ref{eqn:xlags}).

\begin{equation}
    \label{eqn:vort_loc}
    \omega_l\big(x(t)\big)=\frac{m_p K_1 K_2 B}{dz}\bigg(1-e^{-K_1 x(t)}\bigg)
\end{equation} 

By numerical integrating eqn \ref{eqn:vort_simp}, it is found that the Stokes model yields a local vorticity ,$\omega_l\big(x_f\big)$, within $\sim 300\%$ ($\sim 10\%$) of the Kliatcko drag model for the Al(5) particle sizes (Al(1) particle sizes) at our shock conditions. The Cloutman model produces greater vorticity for larger particle sizes as its increased drag coefficient provides greater vorticity production at earlier times. This is apparent in the lag distances calculated for each model, where the Cloutman model provides lower $x_f$, especially at larger particle sizes. While the error for large particles seems more than significant, it is important to note that both drag models show a significantly lower circulation at these sizes.

The total enstrophy equation can be solved by substitution of eqn. \ref{eqn:vort_loc}, along with the Stokes equation for $B_x$, into eqn. \ref{eqn:enst_simp}, applying the divergence theorem to the advection term as before. Integrating this equation and applying the enstrophy (relating it to vorticity), a solution for $\omega_t(t)$ is obtained. The total vorticity produced can then be found as shown in eqn. \ref{eqn:vort_tot}. 

\begin{equation}
    \label{eqn:vort_tot}
    \begin{aligned}
    \omega_t\big(x(t)\big)=& \frac{\sqrt{2} m_p K_1 K_2 B }{dz} \bigg(K_1 x(t) + e^{-K_1 x(t)} - 1 \bigg)^{1/2}
    \end{aligned}
\end{equation}

The maximum value for vorticity can be found for the small-particle limit by taking $\lim_{x_f\to 0} \omega\big(x_f\big)$ (note $\omega_l\big(x(t)\big)$ converges to the same value). At this limit, the maximum vorticity is $\omega_{m}=K_1^2 K_2 m_p u_0 / dz$. Using this maximum, we can non-dimensionalize eqns. \ref{eqn:vort_loc} and \ref{eqn:vort_tot} and solve as a function of $x_s$. 

Up to this point, the solutions are free of interface geometry, considering only a generic interface with length scale $L_0$. At the limit of very small particles ($x_s \to 0$), the vorticity should converge to that of a RMI. For a circular heavy-gas RMI interface, the Picone and Boris \cite{picone1988vorticity} model (eqn \ref{eqn:PBcirc}) provides an estimate for circulation. Since $\Gamma=\int_A \omega dA$, it follows that it will scale equally with the vorticity. 

However, the solutions still depend on a characteristic length scale, $K_1=1/(2L_0)$. The exponential term in eqn \ref{eqn:vort_tot} and \ref{eqn:vort_loc}, $-K_1 x(t)$, implies a length scale for alignment between the particle source term and the deposited vorticity. As the vortcity source term reaches its peak at the midpoint of the circular interface, where the interface tangent is perpendicular to the shock wave, deposited vorticity must travel a distance of $\approx D_0/2$ to fully exit the particle field. Using this length scale, $D_0/2$, for alignment provides $K_1=2/D_0$. With this length scale, and the scaling above, the total circulation produced and the circulation local to the interface can be found as shown in eqns \ref{eqn:ndcirc_tot} and \ref{eqn:ndcirc_loc}.

\begin{equation}
    \label{eqn:ndcirc_tot}
    \begin{aligned}
    \Gamma_{t}=& \sqrt{2} B \bigg(\frac{D_0}{2}\bigg)^2 ln\bigg(\frac{\rho_e}{\rho_B'} \bigg) \bigg( \frac{v_A'}{w_i}-2\bigg) *\bigg(\frac{2}{D_0} x_f + e^{-2x_s/D_0} - 1 \bigg)^{1/2}
    \end{aligned}
\end{equation}

\begin{equation}
    \label{eqn:ndcirc_loc}
    \Gamma_{l}=B \bigg(\frac{D_0}{2}\bigg)^2 ln\bigg(\frac{\rho_e}{\rho_B'} \bigg) \bigg( \frac{v_A'}{w_i}-2\bigg) \big(1-e^{-2x_s/D_0}\big)
\end{equation} 

To examine cases with varying initial circulation deposition (e.g. shock strength and $A_e$) we non-dimensionalize eqns \ref{eqn:ndcirc_tot} and \ref{eqn:ndcirc_loc} as $\Gamma/\Gamma_{PB}$, where $\Gamma_{PB}$ is unique to each experiment or trial. We use the $d_{32}$ value for the characteristic diameter as it represents the mean particle mass to drag area. While the lag distance scales with $d^2$ the momentum carried by each particle scales with $d^3$, thus the $d_{32}$ is the appropriate characteristic diameter.  In order to scale $d_c$ to different particle materials from previous works, we take $d_c$ to be for aluminum particles, scaling other materials as $d_c=d_{32}*\rho_p/\rho_{Al}$ to account for momentum differences due to material density.

\begin{figure}
\includegraphics[width=1\textwidth]{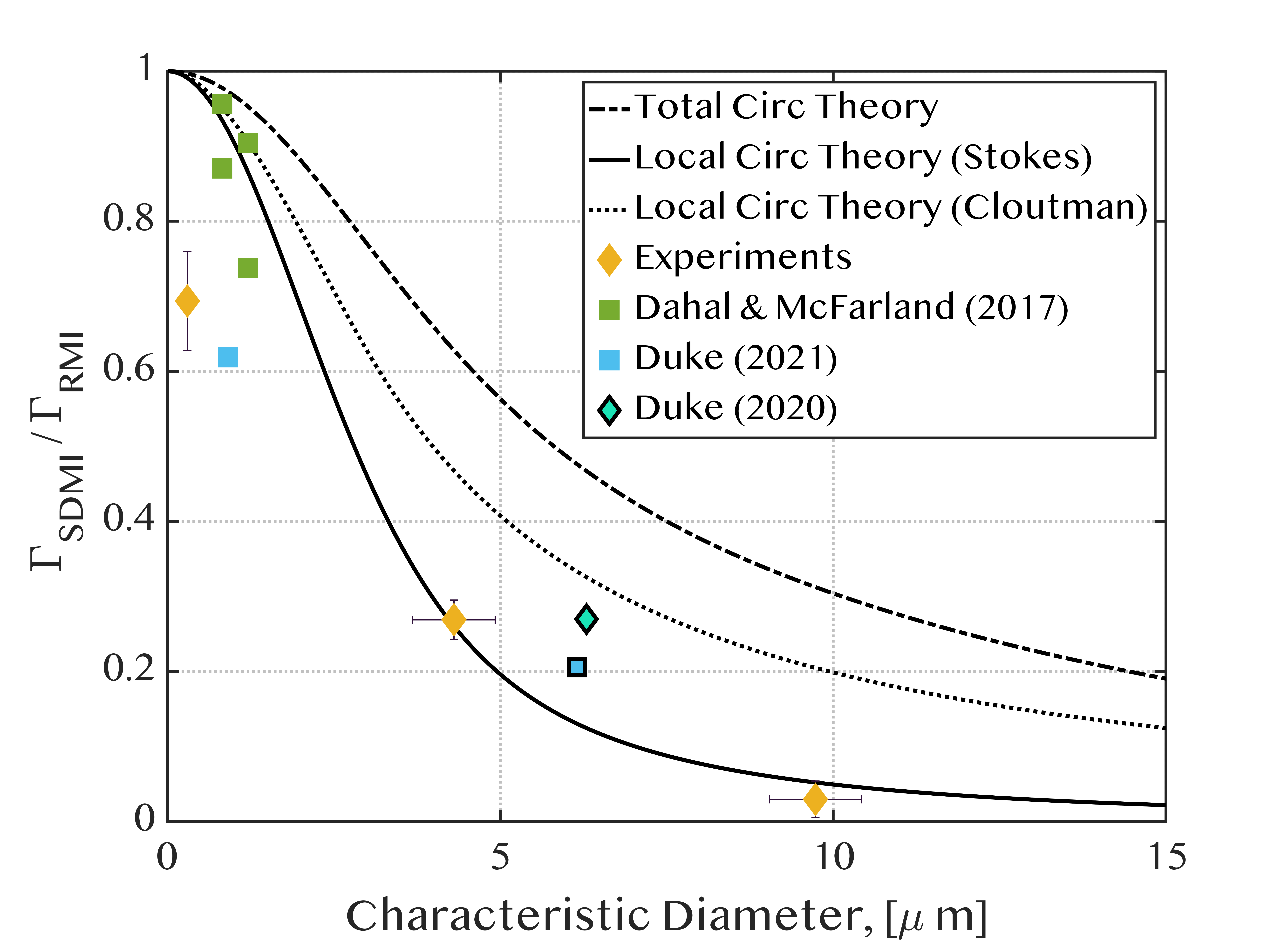}
\caption{\label{fig:circplot} Circulation versus particle characteristic size. Solid Black line: Circulation predicted by local vorticity theory stokes solution. DashedBlack line: Circulation predicted by total vorticity theory. Dotted Black line: Circulation predicted by local vorticity theory cloutman solution. Orange diamonds: Current experimental data. Green squares: Simulations from \cite{dahal2017numerical}. Lightblue squares: Simulations from \cite{duke2021evaporation}. Light Green: Experiments of \cite{duke2020method}. Black outlines indicate experimental data with breakup effects.}
\end{figure}

Figure \ref{fig:circplot} shows the non-dimensionalized results of the equations versus data from the experiments and other SDMI simulations. It can be seen that the local circulation equation agrees better with the measurements. Our experimental methods only measure the circulation of the primary vortex core, excluding the less organized vorticity surrounding the core and missing any vorticity located within unseeded gas. Thus, these measurements represent the local vorticity. The vorticity measurements from simulations follow a similar method, capturing mostly the local vorticty. However, if the vorticity is integrated over the whole domain, the circulation found still agrees best with the local model. The reason for this is likely due to the dissipation of advected vorticty, that is not captured in eqn. \ref{eqn:ndcirc_tot}. Vorticity advected downstream is diffused and dissipated by viscosity. In the case of the simulation results, advected vorticity is dissipated through numerical effects.

The disagreement in circulation for case 1 can be attributed to the finite resolution of the experimental methods. The simulation results of Dahal and McFarland have higher effective resolution and thus match the model better. The small droplet case from the Duke-Walker et al. (2021) simulations shows lower circulation, however. This is attributed in part to the initial water vapor in the interface, which decreases $A_e$, though this alone is likely not enough to explain the discrepancy. For the larger droplet size cases, the Duke-Walker et al. (2021) simulations show greater circulation than predicted by the model but this can be attributed to breakup and evaporation effects which decrease particle lag time \cite{duke2021evaporation,black2017evaporation}. The Duke-Walker et al. (2020) experiments also show greater circulation, due to breakup and evaporation. This experiment also had a significant $A_g$, due to the acetone vapor, that is not captured in the current models.

\section{Summary and Conclusions}
\label{sec:conc}

The effect of particle response time on the Shock-Driven Multiphase Instability (SDMI) was studied in experiments for the first time in the absence of droplet effects (deformation, breakup, evaporation) using well-characterized solid particles. A cylindrical interface seeded with particles was accelerated by a Mach $\sim 1.35$ shock wave leading to hydrodynamic instability and mixing. 

Three cases were studied using different particle sizes measured by Scanning Electron Microscope images ex-situ, and Phase Doppler Particle Anemometry in-situ. In the small-particle limit case, a gas density difference was used with passive tracer particles to create a pure Richtmyer-Meshkov instability (RMI). Cases with increasing particle sizes, from 1 $\mu m$ to 10 $\mu m$, showed the effect of increasing particle relaxation time. In each case the effective gas-particle mixture density was held constant across all cases.

The development of the interface was observed through planar laser Mie-scattering images of the particle field. A reduction in hydrodynamic development with increasing particle size was observed. The small particle case, driven by a gas density difference, followed the development of a classic RMI. The large particle case showed very little development of the primary vortex cores and a long tail of particles lagging behind the interface. 

The vorticity, driving the overall mixing, was measured from the Mie-scattering images using particle imaging velocimetry (PIV). The circulation was then estimated by taking the line-integral of velocity on closed circles around the vortex cores. The maximum vorticity (at a given circle radius) was found for each trial and averaged over ten trials for each case, providing greater statistical certainty in the measurements. The circulation measurements were highly repeatable and showed a strong decrease in circulation with increasing particles sizes. 

A model for the circulation reduction with increasing particle lag distance (size) was proposed based on a RMI circulation deposition model and derivation of theory from the multiphase vorticity equation. Two equations were derived, one for the circulation local to the interface, and one for the total circulation produced, including that which is advected away from the interface. The theory was compared to both the experimental data here, and to previous data from experiments and simulations. The local circulation model is found to agree well with the data from the current experiment and many of the previous results. It is proposed that the diffusion and dissipation of advected vorticity results in only the vorticity local to the interface (the source term) persisting over time. 

This work provides an improved understanding of the mixing process in shock-driven multiphase system. The fundamental theory proposed here can be applied to other interface geometries by selecting appropriate length scales and RMI circulation models. This theory can be applied in applications such as shock-driven combustion and internal confinement fusion to optimize mixing rates. Finally, future work will seek to improve our current model to include gas density gradient effects and competition between baroclinic and multiphase, body-force, vorticity deposition.

\section{Appendixes}

\addcontentsline{toc}{chapter}{BIBLIOGRAPHY}

\bibliographystyle{unsrt}
\bibliography{myreference}   


\end{document}